\documentclass[twocolumns]{aa}
\usepackage{graphicx,longtable,lscape}
\def\ls{{_<\atop^{\sim}}}
\def\gs{{_>\atop^{\sim}}}

\def\mbh{$M_{\rm bh}$}
\def\mbhe{{M_{\rm bh}}}

\def\mstare{{M_{\rm star}}}

\def\msune{{M_{\odot}}}
\def\mdotbh{$\langle \dot{M}_{\rm BH}(M_{\rm BH},z)\rangle$}

\begin{document}
\title{AGN wind scaling relations and the co-evolution of black holes and galaxies }

\author{F. Fiore\inst{1}
\and
C. Feruglio \inst{2}
\and
F. Shankar\inst{3}
\and
M. Bischetti\inst{1}
\and
A. Bongiorno\inst{1}
\and
M. Brusa\inst{4,5}
\and
S. Carniani\inst{6,7}
\and
C. Cicone\inst{8,9}
\and
F. Duras\inst{1}
\and
A. Lamastra\inst{1}
\and
V. Mainieri\inst{10}
\and
A. Marconi\inst{6,11}
\and
N. Menci\inst{1}
\and
R. Maiolino\inst{7}
\and
E. Piconcelli\inst{1}
\and
G. Vietri\inst{1}
\and
L. Zappacosta\inst{1}
}

\institute{
INAF - Osservatorio Astronomico di Roma, via Frascati 33, I00078, Monteporzio Catone, Italy
              \email{fabrizio.fiore@oa-roma.inaf.it}
\and
INAF - Osservatorio Astronomico di Trieste, via G. Tiepolo 11, I-34124 Trieste, Italy 
\and
Department of Physics and Astronomy, University of Southampton, Highfield, SO17 1BJ, UK
\and
Dipartimento di Fisica e Astronomia, Alma Mater Studiorum - Universit\'a di Bologna, viale Berti Pichat 6/2, 40127, Bologna, Italy
\and
INAF - Osservatorio Astronomico di Bologna, via Ranzani 1, 40127 Bologna, Italy 
\and
Dipartimento di Fisica e Astronomia, Universit\'a  di Firenze, Via G. Sansone 1, 50019, Sesto F.no, Firenze, Italy
\and
Cavendish Laboratory, University of Cambridge, 19 J. J. Thomson Ave., Cambridge CB3 0HE, UK
\and
INAF-Osservatorio astronomico di Brera, via Brera 28, 20121, Milan, Italy
\and
ETH, Institute for Astronomy, Department of Physics, Wolfgang-Pauli-Strasse 278093 Zurich, Switzerland
\and
European Southern Observatory, Karl-Schwarzschild-str. 2, 85748 Garching bei M\"unchen, Germany
\and
INAF-Osservatorio Astrofisico di Arcetri, Largo E. Fermi 5, 50125, Firenze, Italy
}

\date{January 21, 2017}

\abstract
{Feedback from accreting supermassive black holes is often identified
  as the main mechanism responsible for regulating star-formation in AGN host
  galaxies. However, the relationships between AGN activity,
  radiation, winds, and star-formation are complex and still far from being
  understood.}
{We study scaling relations between AGN properties, host galaxy
  properties and AGN winds. We then evaluate the
  wind mean impact on the global star-formation history, taking
  into account the short AGN duty cycle with respect to that of
  star-formation. }
{We first collect AGN wind observations for 94
  AGN with detected massive winds at sub-pc to kpc spatial scales.
  We then fold AGN wind scaling relations with AGN luminosity
  functions, to evaluate the average AGN wind mass-loading factor as a
  function of cosmic time.}
{We find strong correlations between the AGN molecular and ionised wind
  mass outflow rates and the AGN bolometric luminosity. The power law
  scaling is steeper for ionised winds (slope 1.29$\pm$0.38) than for
  molecular winds (0.76$\pm$0.06), meaning that the two rates converge
  at high bolometric luminosities.  The molecular gas depletion
  timescale and the molecular gas fraction of galaxies hosting
  powerful AGN driven winds are 3-10 times shorter and smaller than
  those of main-sequence galaxies with similar star-formation rate,
  stellar mass and redshift.  These findings suggest that, at high AGN
  bolometric luminosity, the reduced molecular gas fraction may be due
  to the destruction of molecules by the wind, leading to a larger
  fraction of gas in the atomic ionised phase.  The AGN wind
  mass-loading factor $\eta=\dot M_{OF}$/SFR is systematically higher
  than that of starburst driven winds. }
{Our analysis shows that AGN winds are, on average, powerful enough to
  clean galaxies from their molecular gas only in massive systems at
  z$\ls2$, i.e. a strong form of co-evolution between SMBHs and
  galaxies appears to break down for the least massive galaxies.}
\keywords{Galaxies: active -- Galaxies: evolution -- Galaxies: quasars
  -- general}

\authorrunning {Fiore et et al.}
\titlerunning {AGN wind scaling relations}

\maketitle

%

\section{Introduction}

The past decades have seen a hot debate on whether, and how, the evolution of
galaxies and that of the supermassive black holes (SMBHs) hosted in
their nuclei is correlated.

The debate started with the HST discovery of SMBHs in most local
bulges (Richstone et al. 1998). SMBH mass and host bulge properties --
such as velocity dispersion, luminosity and mass -- were found to
tightly correlate with each other (Gebhardt et al. 2000, Ferrarese \&
Ford 2005, Kormendy \& Ho 2013 and references therein, but also see
Shankar et al. 2016,2017).  Furthermore, the comparison of the SMBH mass
function derived from AGN luminosity function and from the local bulge
luminosity function suggests that SMBH growth is mostly due to
accretion of matter during their active phases, and therefore that
most bulge galaxies passed a phase of strong nuclear activity (Soltan
1982, Marconi et al. 2004, Shankar et al. 2004, Merloni \& Heinz
2008). Both findings seemed to imply links between SMBH accretion and
bulge formation, i.e. a {\it strong} form of AGN/galaxy co-evolution.
Indeed, soon after the discovery of the SMBH-bulge relationships,
several authors (Silk \& Rees 1998, Fabian 1999, King 2003, Granato et
al. 2004) suggested that they can be naturally explained if AGN winds
efficiently interact with the galaxy ISM. When the black hole reaches
a critical mass it may be powerful enough to heat up and eject the gas
from the galaxy, terminating the growth of both SMBH and galaxy, and
giving rise to the observed scaling between SMBH mass and bulge
velocity dispersion.  AGN feedback can not only modify AGN host
galaxies it can also affect the intra-cluster matter (ICM) in groups
and clusters of galaxies.  Two modes for AGN feedback have been indeed
postulated. The so called {\it radio-mode} in the central cluster
galaxies, and the {\it quasar-mode}, characterized by slower winds of
both ionized, neutral atomic, and molecular matter.

Radio-mode feedback is evident in cool core clusters and groups, where
the ICM is heated up by AGN jet-driven radio bubbles. The power to
excavate cavities in the ICM is proportional to the X-ray luminosity,
and the power in cavities is proportional to the AGN radio luminosity
(see McNamara \& Nulsen 2007, Cattaneo et al. 2009, Fabian 2012 for
reviews). Interestingly, only the brightest central galaxies (BCGs) in
clusters/groups with low inner entropy (short cooling time) have an
active nucleus, {\it and} are actively forming stars (Cavagnolo et
al. 2008, 2009).  The situation is best described by Voit \& Donahue
(2015): {\it a delicate feedback mechanism where AGN input energy
  regulates the gas entropy and in turn further gas accretion and
  star-formation (stars can form from low entropy, cold and dense gas
  only)}. Thus a multi-phase gas structure naturally develops in
cluster cores and within the BCGs leading to AGN feedback triggered by
cold accretion (Gaspari et al. 2012, 2013, 2014, 2016).

Similar autoregulation may occur in galaxies other than BCGs, where
feedback might be due to more common AGN winds. Indeed several direct
observation of ISM modifications by AGN winds have been collected so
far. Cano-Diaz et al. (2012), Cresci et al.  (2015), and Carniani et
al. (2016) found that AGN winds and actively star-forming regions are
spatially anti-correlated.  Similarly, Davies et al. (2007) and Lipari
et al. (2009) found little evidence for young (Myrs) stellar
populations in the $\ls1$ kpc region of Markarian 231, where a
powerful molecular outflow is observed (Feruglio et al. 2010, 2015).
However, although promising, these quasar-mode feedback observations
are still too sparse to derive strong conclusions.

The correlation between SMBHs and bulge properties do not necessarily
require feedback, and can be also explained if SMBHs and bulges formed
simultaneously, during episodes when a fixed fraction of gas accretes
toward the central black hole while the rest forms the spheroid stars.
Menci et al. (2003) reproduced the BH mass -- $\sigma_{bulge}$
correlation as the combination of three factors: a) the merging
histories of the galactic dark matter clumps, implying that the mass
of the available cold gas scales as $\sigma^{2.5}$; b) the
destabilisation of cold gas by galaxy interactions, which steepens the
correlation by another factor $\sigma$; and c) SNe feedback, which
depletes the residual gas content of shallow potential wells, further
steepening the correlation.  Later, Peng (2007) showed that galaxy
mergers are efficient in averaging out extreme values of
M$_{BH}$/M$_*$, converging toward a narrow correlation between these
quantities, close to the observed one, even starting from arbitrary
distributions. Jahnke \& Macci\'o (2011) showed that the number of
mergers needed to this purpose is consistent with that of standard
merger tree models of hierarchical galaxy (and SMBH) formation.  In
this scenario the SMBHs and bulges do not necessarily know about each
other. No causal connection exists between these systems, and their
properties are connected just by natural scaling relations. We can
call this as a {\it weak} form of AGN/galaxy co-evolution.  More
recently, the analysis of Shankar et al. (2016) supports a strong
dependence between SMBH mass and bulge velocity dispersion, while the
dependence with the bulge mass is weaker, disfavouring this scenario,
and suggesting to investigate AGN/galaxy co-evolution independently
from the SMBH mass -- bulge mass scaling relations.

Comparing model predictions to the observed SMBH mass -- bulge
properties hardly allows one to discriminate between {\it weak} and
{\it strong} forms of AGN/galaxy co-evolution. This is probably due to
the fact that SMBH mass and bulge properties are quantities {\it
  integrated} along cosmic time, with SMBHs and bulges assembled
during the Hubble time, as a consequence of several merging and
accretion events. A different route attempted to distinguish between
{\it weak} and {\it strong} forms of co-evolution, is to study
derivative quantities, such as the SMBH accretion rate and the
star-formation rate (SFR), or, the cosmological evolution of the AGN
and galaxy luminosity densities.  Franceschini et al. (1999) were
among the first to realise that the luminosity dependent evolution of
AGN, with lower luminosity AGN peaking at a redshift lower than
luminous QSOs (Ueda et al. 2003, Fiore et al. 2003, La Franca et
al. 2005, Brandt \& Hasinger 2005, Bongiorno et al. 2007, Ueda et
al. 2014, Aird et al. 2015, Brandt \& Alexander 2015), mirrors that of
star-forming galaxies and of massive spheroids. These trends, dubbed
``downsizing'' by Cowie et al. (1996), and in general the relationship
between the evolution of AGN and galaxy growth, may arise from
feedback mechanisms linking nuclear and galactic processes.

Indirect evidence for AGN feedback come from the statistical
properties of AGN host galaxies with respect to the inactive
population. It is well known since the pioneering HST studies of
Bahcall et al. (1997) that luminous QSOs reside preferentially in
massive, spheroid-dominated host galaxies, whereas lower luminosity
QSOs are found in both spheroidal and disky galaxies (Dunlop et
al. 2003, Jahnke et al. 2004 and references therein). The distribution
of AGN host galaxy colors, morphologies, SFR, specific SFR are wider
than that of star-forming galaxies of similar masses, and skewed
toward redder/more inactive galaxies (e.g. Alexander et al. 2002,
Mignoli et al. 2004, Brusa et al. 2005, Nandra et al. 2007, Brusa et
al. 2009, 2010, Mainieri et al. 2011, Bongiorno et al. 2012,
Georgakakis et al. 2014). Many AGN are hosted in red-and-dead
galaxies, or lie in the so called green valley.  Recent ALMA
observations of X-ray selected AGN in the GOODS field (Mullaney et
al. 2015) confirmed these earlier results, showing that the bulk of
the AGN population lie below the {\it galaxy main-sequence}, (see
Daddi et al. 2007, Rodighiero et al. 2011, and refs. therein). Because
the stellar mass function of star-forming galaxies is exponentially
cut-offed above a quenching mass M$_*\sim10^{11}$ M$_\odot$ (Peng et
al. 2010), the galaxy main-sequence flattens above the same mass,
whereas the star-formation efficiency and the gas-to-star mass
fraction decrease (Genzel et al. 2010 and references therein). AGN
feedback may well be one of the drivers of these transformations, as
well as the main driver for the quenching of star-formation in massive
galaxies (Bongiorno et al. 2016), pointing toward a {\it strong} form
of AGN/galaxy co-evolution. We explore this possibility in this paper.

This paper is organized as follows. In section 2 we review AGN massive
wind observations, and study the scaling relationships between wind
mass outflow rate, velocity, kinetic power, momentum load, AGN
bolometric luminosity and host galaxy SFR. We then plug AGN wind
studies in the broader scenario of star-forming galaxies scaling
relations (Genzel et al. 2015 and references therein), to understand
whether AGN hosting strong winds are outliers in these
relationships. We study the relationships between the depletion
timescale (the ratio between molecular gas mass and SFR), and gas
fraction (the ratio between molecular gas mass and galaxy stellar
mass), with the offset from the galaxy main sequence, redshift and
host galaxy stellar mass, for a sample of sources with interferometric
molecular measurements.  In section 3 we evaluate the wind statistical
relevance on the global star-formation history, by folding the AGN
wind scaling relations with the AGN luminosity functions. This allows
to account for the fact that AGN shine in a relatively small fraction
of galaxies, i.e. the AGN timescales are usually shorter than the
star-formation timescales. We compare the cosmic, average AGN outflow
rate, computed by using the AGN wind scaling relations, to the galaxy
cosmic star-formation rate, to study the regimes (galaxy masses,
cosmic epoch) where AGN winds are statistically strong enough to
affect star-formation in the global galaxy population.  Section 4
presents our conclusions.  A $H_0=70$ km s$^{-1}$ Mpc$^{-1}$,
$\Omega_M$=0.3, $\Omega_{\Lambda}=0.7$ cosmology is adopted
throughout.

\section{AGN wind scaling relations}

Although wind observations are very common in AGN (see Elvis 2000,
Veilleux et al. 2005 and Fabian 2012 for reviews), most studies
concern ionised gas and uncertain spatial scales. In the past few
years the situation changed drastically. Several fast ($v_{OF}$ of the
order of 1000 km/s), massive outflows of ionised, neutral and
molecular gas, extended on kpc scales, have been discovered thanks to
three techniques:
1) deep optical/NIR spectroscopy, mainly from integral field
observations (IFU, e.g. Nesvadba et al. 2006,2008; Alexander et
al. 2010; Rupke \& Veilleux 2011; Riffel \& Storchi-Bergmann 2011;
Cano-Diaz et al. 2012; Greene et al. 2012, Harrison et al. 2012, 2014;
Liu et al. 2013a,b; Cimatti et al. 2013, Tadhunter et al. 2014; Genzel
et al. 2014; Brusa et al. 2015a, Cresci et al. 2015; Carniani et
al. 2015; Perna et al. 2015a,b, Zakamska et al. 2016);
2) interferometric observations in the (sub)millmetre domain
(e.g. Feruglio et al. 2010,2013a,b, 2015; Alatalo et al. 2011, Aalto
et al. 2012, Cicone et al. 2012, 2014, 2015; Maiolino et al. 2012,
Krips et al. 2011, Morganti et al. 2013a,b, Combes et al. 2013,
Garcia-Burrillo et al. 2014); and
3) far-infrared spectroscopy from Herschel (e.g. Fischer et al. 2010,
Sturm et al. 2011, Veilleux et al 2013, Spoon et al. 2013, Stone et al. 2016,
Gonzalez-Alfonso et al. 2016).  
In addition, AGN-driven winds from the accretion disc scale up to the
dusty torus are now detected routinely both in the local and in the
distant Universe, as blue-shifted absorption lines in the X-ray
spectra of a substantial fraction of AGN (e.g. Piconcelli et al. 2005,
Kaastra et al. 2014).  The most powerful of these winds, observed in
20-40\% of local AGN (e.g. Tombesi et al. 2010) and in a handful of
higher redshift objects (e.g. Chartas et al. 2009, Lanzuisi et
al. 2012), have extreme velocities (Ultra-Fast Outflows, UFOs,
v~0.1-0.3c) and are made by highly ionized gas which can be detected
only at X-ray energies.

We collected from the literature observations of AGN with reliable
massive outflow detections, for which there is an estimate (or a
robust limit) on the physical size of the high velocity gas involved
in the wind. The sample includes molecular winds, ionised winds (from
[OIII], H$\alpha$ and H$\beta$ lines), broad absorption line (BAL)
winds and X-ray absorbers (both UFOs and the slower ``warm
absorbers''). We give in Appendix A a short description of the source
samples used in the following analysis.

\begin{figure*}
\begin{tabular}{cc}
\includegraphics[width=8.5cm]{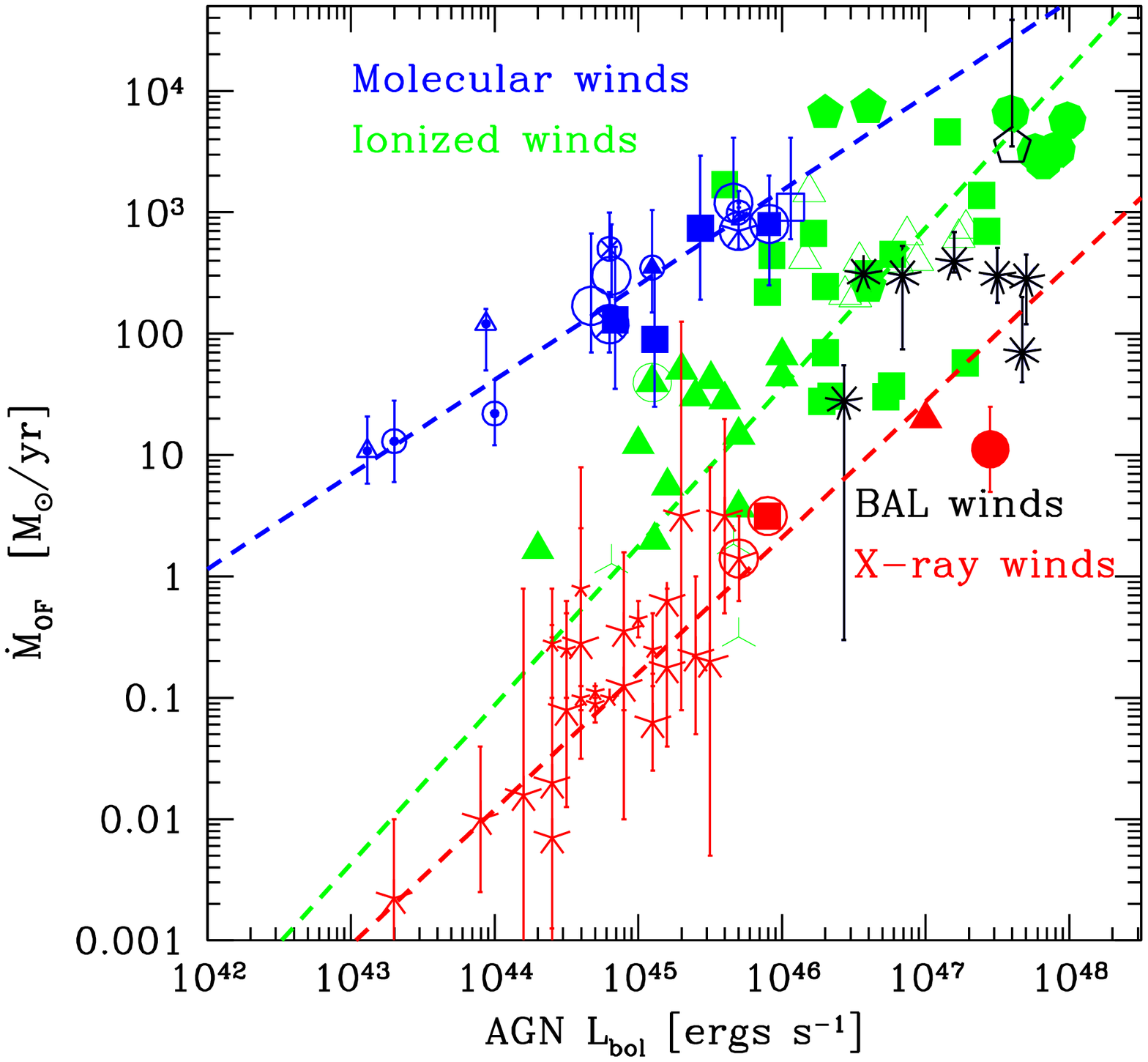}
\includegraphics[width=8.5cm]{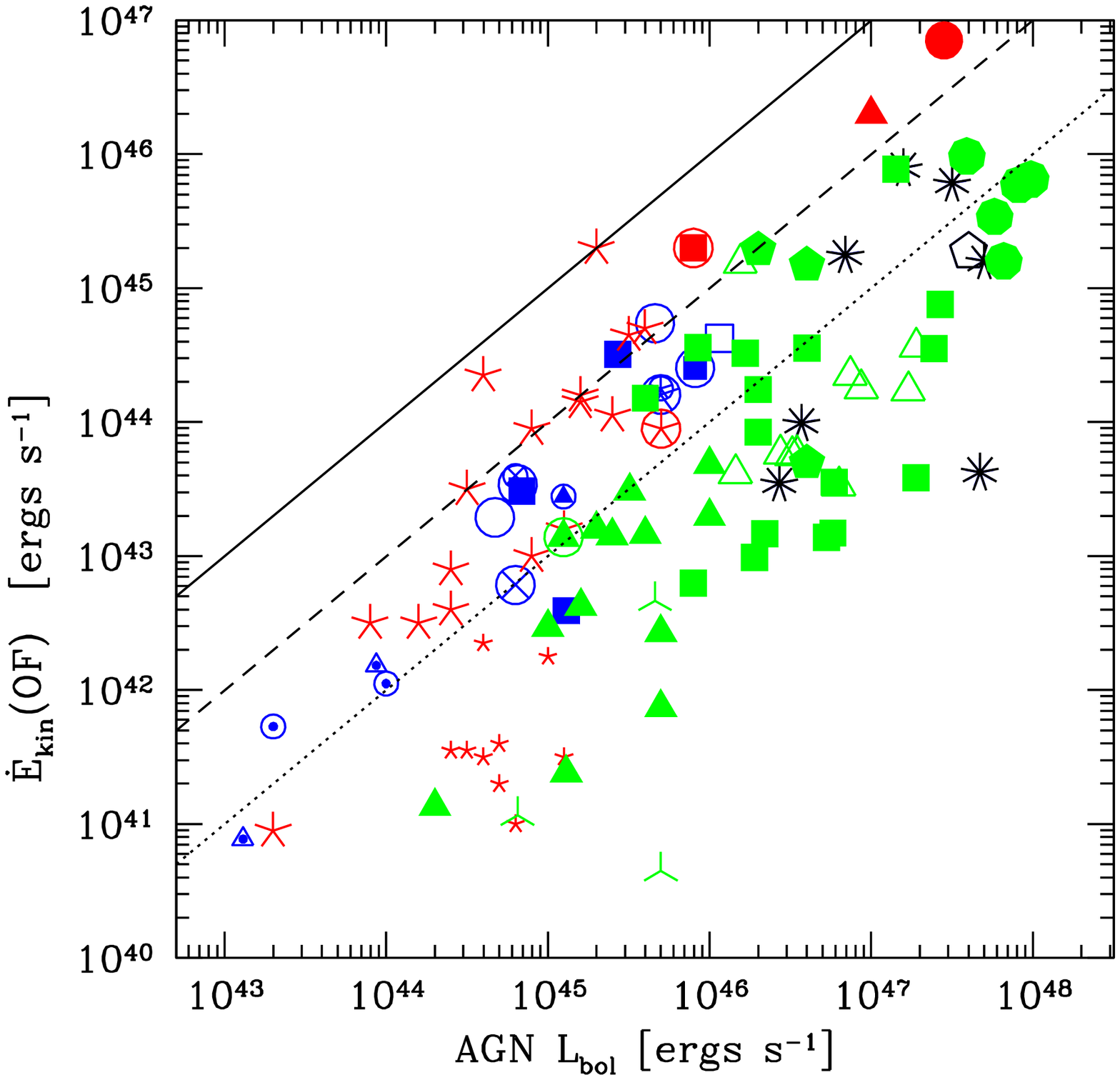}
\end{tabular}
\caption{ [Left panel]: The wind mass outflow rate as a function of
  the AGN bolometric luminosity.  AGN for which molecular winds have
  been reported in the literature (mostly local ULIRGs and Seyfert
  galaxies) are shown with blue symbols. In particular: open circles
  are CO outflows; the open square is the measurement for
  IRAS23060; filled squares are OH outflows; the starred open
  circles are for Markarian 231 (large symbol for the outflow measured
  within $R_{OF}=1$ kpc and small symbol for the outflow at
  $R_{OF}=0.3$ kpc); the crossed open circles are the measurements for
  NGC6240 (large symbol for $R_{OF}=3.5$ kpc and small symbol for
  $R_{OF}=0.6$ kpc); the small dotted open triangle marks the
  measurement in the Circum Nuclear Disk of NGC1068 ($R_{OF}=0.1$ kpc)
  and NGC1433 ($R_{OF}=0.06$ kpc); the small dotted open circles
  represent the measurements for NGC1266, IC5066 at $R_{OF}=0.5$ kpc;
  the squared open circle marks IRASF11119+13257 measurement at
  $R_{OF}=0.3$ kpc.  Green symbols mark ionised outflows
  measurements. In details: filled squares mark z$>$1 AGN; filled
  triangles mark z=0.1-0.2 AGN; open triangles mark z=0.4-0.6 type 2
  AGN; pentagons mark z=2-3 radiogalaxies; filled circles mark
  hyper-luminous z=2-3 QSOs.  BAL winds are shown with black stars.
  The black open pentagon highlights the [CII] wind in J1148+5251 at
  z=6.4.  Finally, red symbols mark X-ray outflows. In details:
  large five pointed stars are local UFOs; the starred open circle,
  the filled triangle and the circled square are the measurement for
  Markarian 231, PDS456 and IRASF11119+13257, respectively. Small five
  point stars are slower warm absorbers.  The dashed blue, green and
  red lines are the best fit correlations of the molecular, ionised,
  and X-ray absorber samples, respectively.  [Right panel]: Wind
  kinetic power as a function of the AGN bolometric luminosity.
  Solid, dashed and dotted line represent the correlations $\dot
  E_{kin}=1,0.1,0.01 L_{bol}$.  }
\label{lbolmdot}
\end{figure*}

We have recomputed the wind physical properties (mass outflow rate,
kinetic energy rate) using the same assumptions for all sources of
each sample (as detailed in Appendix B). While wind geometry, wind gas
density, temperature, metallicity etc. may well differ from source to
source, applying a uniform analysis strategy minimizes systematic
differences from sample to sample. In fact, self-consistent
information of the gas physical and chemical properties is not
available for the majority of the sources with detected winds, and
thus assumptions on these properties must be done in any case. For
ionized wind parameters, the chain of assumptions needed to convert
observed quantities into physical quantities is particularly long (see
Appendix B), and therefore the largest uncertainties concern these
winds (about one order of magnitude or even more, see Harrison et
al. 2014).  We also collected from the literature AGN and galaxy
properties, such as luminosities, SFRs, stellar masses, molecular gas
masses. We note that these quantities are calculated by different
authors, using non-homogeneous recipes. In particular, bolometric
luminosity are calculated either from fitting optical-UV spectral
energy distributions (SEDs) with AGN templates and from X-ray or
infrared luminosities by applying a bolometric correction. Most SFRs
are calculated from far infrared luminosities and therefore are not
instantaneous SFRs. Stellar masses are calculated from modelling
optical-near-infrared galaxy SEDs with galaxy templates or by
converting near infrared luminosities from IFU observations of nearby
AGN host galaxies into stellar masses. Molecular gas masses are
calculated converting CO luminosities into H$_2$ gas masses, by
assuming a standard conversion factor (see Appendix for details).
This unavoidably introduces some scatter in the correlations discussed
in the following sections.

Altogether, we have assembled a sample of 109 wind measurements of 94
AGN with detected massive winds at different scales (sub-pc to kpc)
and ionisation states, that we use to constrain the relationships
between wind parameters, AGN parameters and host galaxy
parameters. This sample is definitely not complete and suffers from
strong selection biases; above all, we note that most molecular winds
and UFOs are found in local ULIRGs and Seyfert galaxies. Ionised winds
are found in both low-redshift AGN and z=2-3 luminous/hyper-luminous
QSOs. BALs are from z=2-3 QSOs.

\subsection{Wind parameters vs. AGN parameters}

\begin{figure*}
\begin{tabular}{cc}
\includegraphics[width=8.5cm]{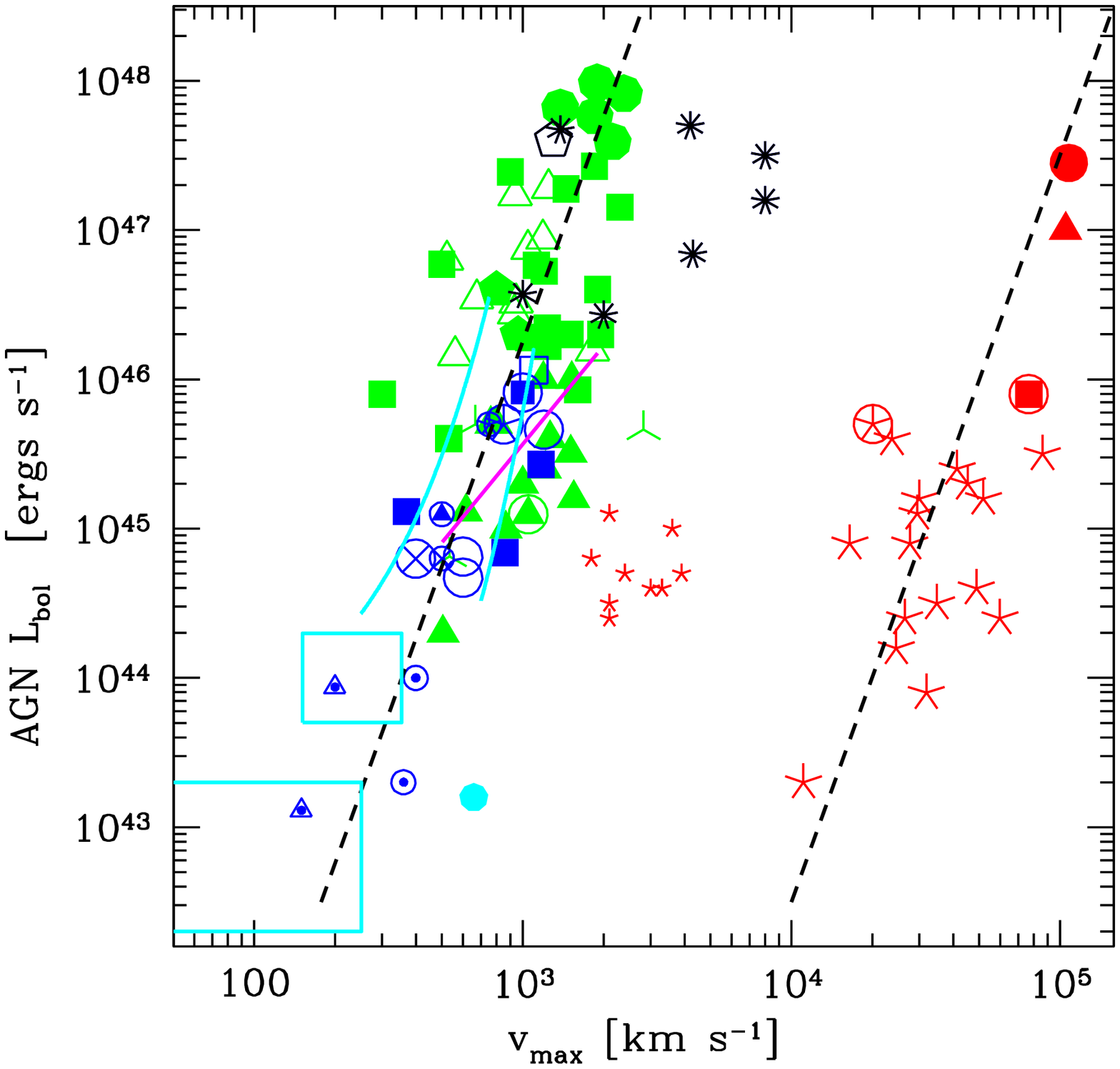}
\includegraphics[width=8.5cm]{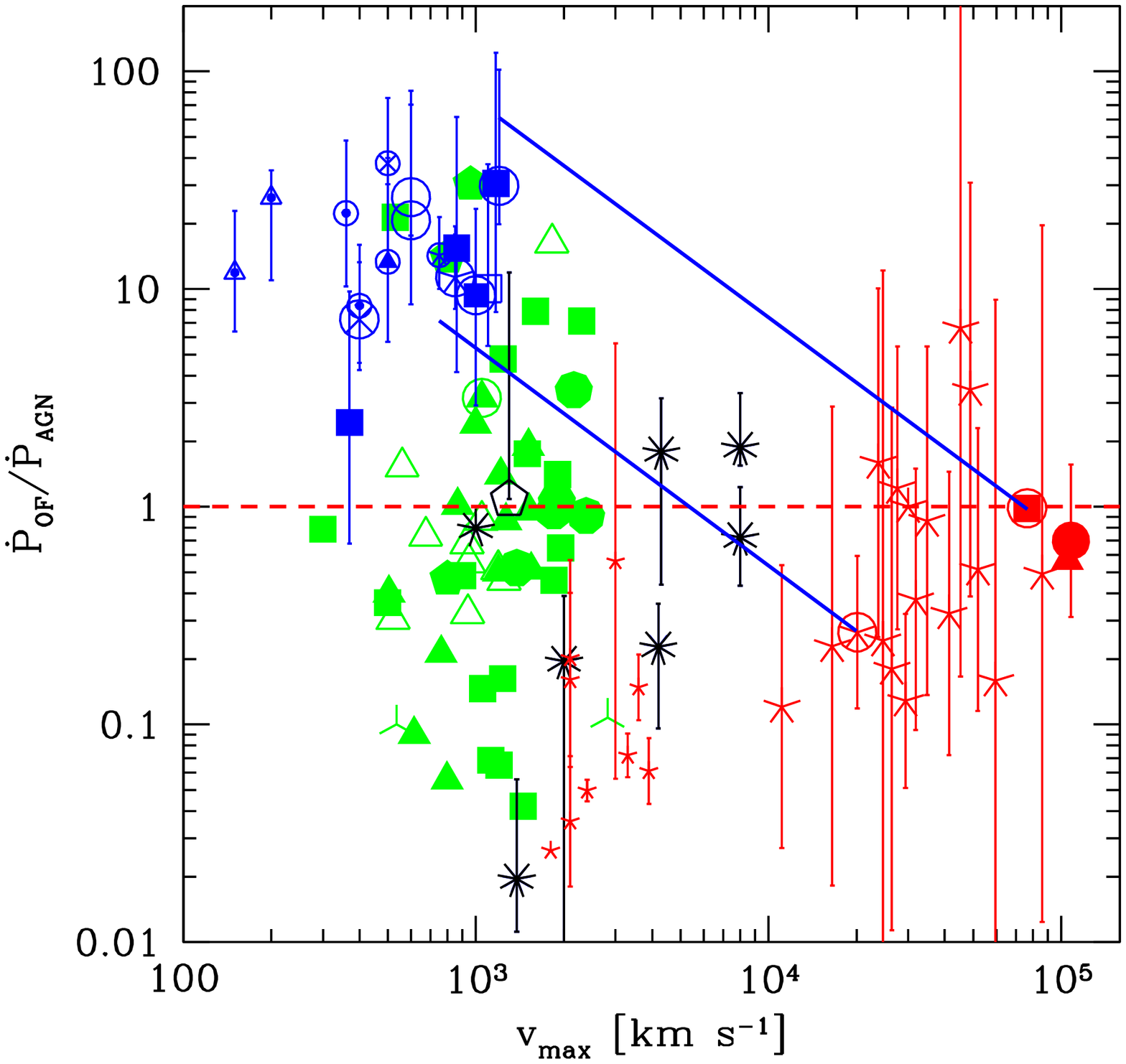}
\end{tabular}
\caption{ [Left panel]: AGN bolometric luminosity as a function of the
  maximum wind velocity, $v_{max}$. The black dashed lines mark a
  $v_{max}^5$ scaling. The magenta solid line is the best fit
  correlation found by Spoon et al. 2013 for OH outflows. The two cyan
  solid lines are the best fit scaling found by Veilleux et al. 2013
  for OH outflows, using $v_{max}$ and $v_{80}$. The cyan boxes and
  filled dot are the loci covered by two groups of Swift BAT AGN with
  $42.3<L_{bol}<43.3$ and $43.3<L_{bol}<43.3$ and by the outlier
  NGC7479, from Stone et al. (2016). [Right panel]: Wind momentum load
  (outflow momentum rate divided by the AGN radiation momentum rate
  L/c) as a function of $v_{max}$. The red dashed line mark the
  expectations for a momentum conserving outflow. The two blue solid
  lines mark the expectations for pure energy conserving outflows for
  Markarian 231 (starred circle) and IRASF11119+13257 (squared
  circle). Symbols as in Fig. \ref{lbolmdot}.  }
\label{pload}
\end{figure*}

Fig. \ref{lbolmdot} shows the wind mass outflow rate (left panel) and
kinetic power (right panel) as a function of the AGN bolometric
luminosity. The mass outflow rate and kinetic power of molecular winds
(blue symbols) are correlated rather well with the AGN bolometric
luminosity (see Table 1, which gives for each correlation the Spearman
rank, SR, correlation coefficient, the probability of the correlation
and the best fit slope, obtained from a least square fit between the
two variables). The log linear slope is $0.76\pm0.06$ for the mass
outflow rate and $1.27\pm0.04$ for the kinetic power.  
The average ratio $\dot E_{kin}/L_{bol}$ in the molecular winds sample
is 2.5\%.  

Ionised winds (green symbols), BAL winds (black symbols),
and X-ray absorbers (red symbols), lie below the correlation found for
molecular winds. Most ionised winds have $\dot M_{OF}$ 10-100 times
smaller than molecular winds at L$_{bol}\ls10^{46}$ ergs/s. Above this
luminosity, ionised winds have $\dot M_{OF}$ similar or a few times
lower than molecular winds. There is a good correlation between $\dot
M_{OF}$, $\dot E_{kin}$, and the bolometric luminosity for ionised
winds (see Table 1) with log linear slopes $1.29\pm0.38$ and
1.50$\pm0.34$ respectively.  The average ratio $\dot E_{kin}/L_{bol}$
for the ionised winds sample is 0.16\% at logL$_{bol}=45$ and 0.30\%
at logL$_{bol}=47$.

X-ray absorbers and BAL winds have respectively $\dot M_{OF}\sim500,
30$ times lower than what expected from the best fit linear
correlation for molecular winds, again showing a trend for higher
differences with respect to molecular winds at lower bolometric
luminosities.  About half X-ray absorbers and BAL winds have $\dot
E_{kin}/L_{bol}$ in the range 1-10\% with another half having $\dot
E_{kin}/L_{bol}<$ 1\%.

The left panel of Fig. \ref{pload} show the AGN bolometric luminosity
as a function of the maximum wind velocity, $v_{max}$, defined
following Rupke \& Veilleux (2013) as the shift between the velocity
peak of broad emission lines and the systemic velocity plus 2 times
the $\sigma$ of the broad gaussian component, see the Appendix.
$v_{max}$ correlates with the bolometric luminosity for molecular
winds, and ionised winds. Considering the two winds together again
produces a strong correlation and a log linear slope of 4.6$\pm$1.5
(see Table 1).  For X-ray absorbers the situation is more complex,
since they are divided in two broad groups, warm absorbers with lower
velocities and UFOs with higher velocities. For UFOs with
$v_{max}>10^4$km/s the correlation between AGN bolometric luminosity
and maximum velocity is still remarkably strong, with a log linear
slope of 3.9$\pm$1.3 (Table 1), statistically consistent with that of
molecular+ionized winds. This means that at each given bolometric
luminosity the ratio between UFO maximum velocity and
molecular-ionized wind maximum velocity is similar, and equal to
$\sim40-50$. We also report in Fig. \ref{pload} the scalings found by
Spoon et al. (2013) and Veilleux et al. (2013) for OH outflows in
samples of ULIRGs and QSOs at z$<$0.3. Four of the objects in Veilleux
et al. 2013 are also part of our sample, see Table B1.

BALs and the lower velocity X-ray absorbers $v_{max}<10^4$ km/s (the
so called X-ray warm absorbers), also seem to show a correlation
between AGN bolometric luminosity and maximum velocity, with a slope
close to the fourth-fifth power, with the warm absorbers present in
low luminosity systems and BALs present in high luminosity systems.

The right panel of Fig. \ref{pload} shows the wind momentum load
(i.e. the wind momentum rate, $\dot P_{OF}=\dot M_{OF}\times v_{max}$,
divided by the AGN radiation momentum rate, $\dot P_{AGN}=L_{bol}/c$)
as a function of $v_{max}$ (see also Stern et al. 2016). The blue
solid lines are the expectations for energy conserving winds ($\dot
P_{OF}/\dot P_{AGN}\approx v_{UFO}/v_{OF}$) for the cases of Markarian
231 and IRASF11119+13257, the only two sources for which both X-ray
winds and molecular winds have been detected (Tombesi et al. 2015,
Feruglio et al. 2015).  Molecular winds have momentum load in the
range 3-100, about half have momentum load $>10$, suggesting again
that most massive-extended outflows are not momentum conserving but
rather energy conserving winds, extended on the host galaxy scales.

Ionised winds have velocities intermediate between
molecular winds and X-ray absorbers. The range of their momentum load
is wide, from 0.01 to 30.  Most BAL and X-ray winds have $\dot
P_{OF}/\dot P_{AGN}\ls1$, suggesting that they are probably momentum
conserving, as predicted by the King (2003) model.

\begin{figure*}
\centering
\begin{tabular}{cc}
\includegraphics[width=8.5cm]{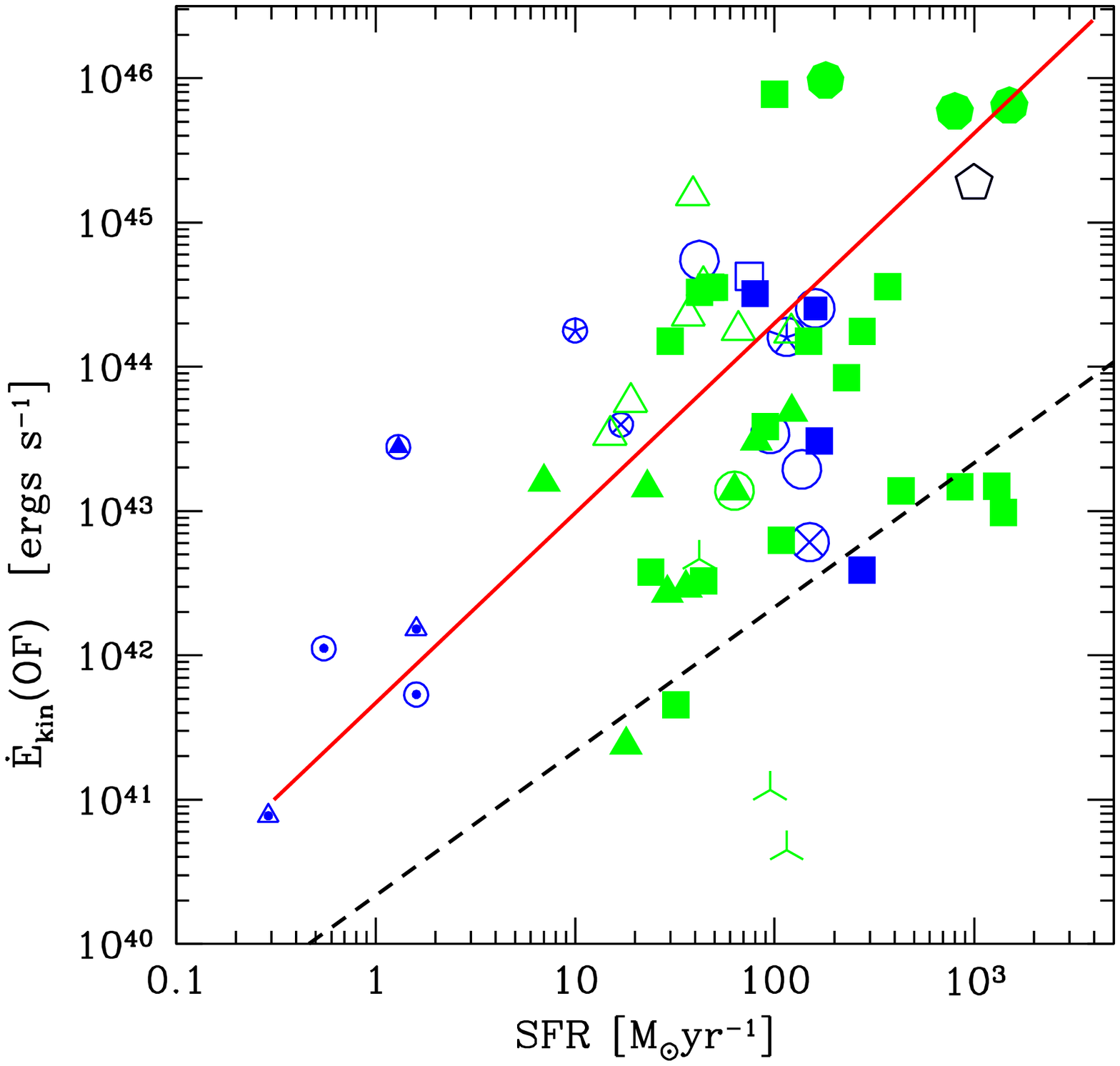}
\includegraphics[width=8.5cm]{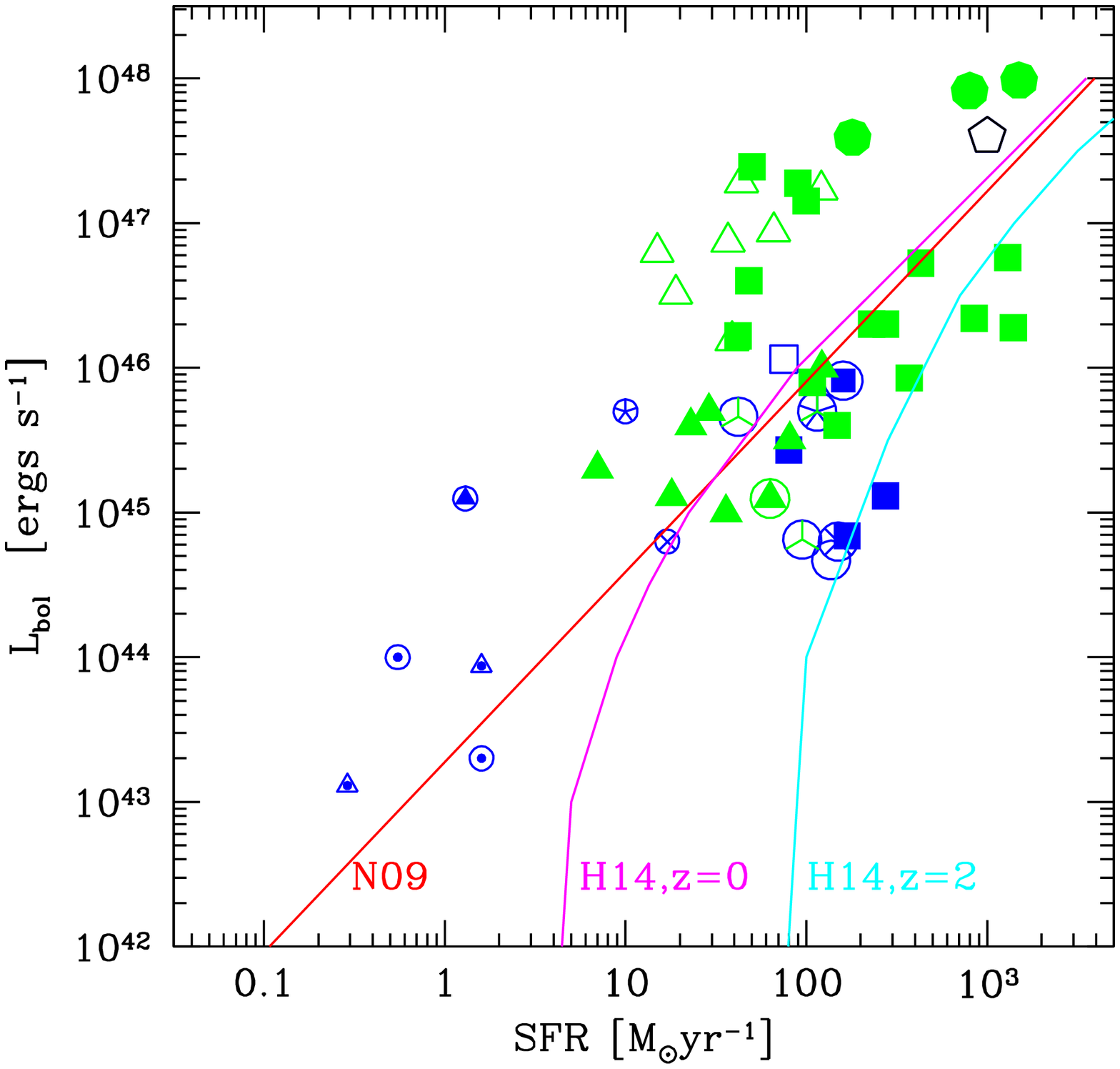}
\end{tabular}
\caption{[Left panel:] Outflow kinetic power as a function of the
  star-formation rate in the host galaxy (computed, when possible, in
  a region similar to that where the outflow has been detected). The
  dashed line is the expectation of a SN-driven wind, by assuming
  0.0066 SNe per solar mass of newly formed star (Salpeter IMF) a
  total luminosity for each SN of $10^{51}$ ergs/s and an efficiency
  of releasing this luminosity in the ISM to drive a shock of 10\%.
  The solid red line is the expected SFR obtained using the Netzer
  (2009) relationship between SFR and AGN bolometric luminosity and
  assuming the average $\dot E_{kin}/L_{bol}=0.025$ found for
  molecular winds in Fig. \ref{lbolmdot}. [Right panel:] AGN
  bolometric luminosity as a function of the host galaxy
  star-formation rate.  The red, magenta and cyan lines in the right
  panel are the expected relations based on the SFR$-L_{bol}$
  correlations by Netzer (2009), Hickox et al. (2014; z=0) and Hickox
  et al. (2014, z=2), respectively. Symbols as in
  Fig. \ref{lbolmdot}.}
\label{ekinsfr}%
\end{figure*}

\begin{table*}
\footnotesize
\begin{minipage}[!h]{1\linewidth}
\centering
\caption{Correlations of wind parameters with AGN bolometric luminosity and host galaxies SFR}
\begin{tabular}{lcccc} 
\hline
Correlation      & slope &  Spearman Rank (SR) &  d.o.f. & null hypothesis probability \\
\hline
\multicolumn{5}{c}{Molecular outflows} \\
$\dot M_{OF}$ vs. L$_{bol}$   & 0.76$\pm0.06$ &  0.86   & 15  & $<10^{-5}$ \\
$\dot E_{kin}$ vs. L$_{bol}$   & 1.29$\pm0.08$ &  0.87  & 15  & $<10^{-5}$\\ 
v$_{max}$ vs. L$_{bol}$ & 3.4$\pm$0.5 & 0.80 & 15  & $<10^{-5}$\\
L$_{bol}$ vs SFR & 0.9$\pm$0.3 &  0.45 & 15 & 3.6\% \\
$\dot E_{kin}$ vs SFR & 1.2$\pm$0.4 & 0.33 & 15 & 4.3\% \\
\hline
\multicolumn{5}{c}{Ionised outflows} \\
$\dot M_{OF}$ vs. L$_{bol}$    & 1.29$\pm0.38$ & 0.72 & 49 & $<10^{-5}$\\  
$\dot E_{kin}$ vs. L$_{\rm bol}$   & 1.48$\pm0.37$ &  0.75 & 49 & $<10^{-5}$\\ 
v$_{max}$ vs. L$_{bol}$ &  6.1$\pm$4.4 & 0.34  & 49 & 0.7\%\\
L$_{bol}$ vs SFR & 2.2$\pm$1.6 & 0.34 & 34 &  2\%\\
$\dot E_{kin}$ vs SFR & 4.9$\pm$4.4 & 0.19  & 34 & 13\% \\
\hline
\multicolumn{5}{c}{Xray outflows} \\
$\dot M_{OF}$ vs. L$_{bol}$  & 1.12$\pm$0.16 & 0.75 & 27 & $<10^{-5}$ \\  
$\dot E_{kin}$ vs. L$_{bol}$  & 2.0$\pm0.4$ & 0.69 & 27 &  $<10^{-5}$\\ 
\hline
\multicolumn{5}{c}{Molecular + Ionised outflows} \\
v$_{max}$ vs. L$_{bol}$ & 4.6$\pm$1.8 & 0.57 & 67  &  $<10^{-5}$\\\
L$_{bol}$ vs SFR & 1.5$\pm$0.6 & 0.44  & 52 &  $4\times10^{-4}$\\
$\dot E_{kin}$ vs SFR & 2.1$\pm$1.4 & 0.28   & 52 & 2\% \\
\hline
\multicolumn{5}{c}{Ultra Fast Outflows} \\
$\dot M_{OF}$ vs. L$_{bol}$  & 1.13$\pm$0.11 & 0.90 & 18 & $<10^{-5}$ \\  
$\dot E_{kin}$ vs. L$_{bol}$  & 1.44$\pm$0.11 & 0.87 & 18 &  $<10^{-5}$\\ 
v$_{max}$ vs. L$_{bol}$  & 3.9$\pm$1.4 & 0.46 & 18 & 0.4\% \\

\hline
\hline
\end{tabular}
\end{minipage}
\end{table*}

\subsection{Wind parameters vs. host galaxy star-formation rate}

We now study the correlations between massive, extended winds, i.e. molecular and
ionised winds, and the properties of their host galaxies.

Figure \ref{ekinsfr} shows the outflow kinetic power and AGN
bolometric luminosity as a function of SFR in the host galaxy
(correlation coefficients given again in Table 1). There is a loose
correlation between log($\dot E_{kin}$) and log(SFR).  It should be
kept in mind that the SFR plotted in Fig. \ref{ekinsfr} is, in most cases, not the
instantaneous SFR but rather the conversion from the observed
FIR luminosity.  The instantaneous SFR can be zero in these systems,
and what we are observing is light from stars born hundreds of
millions of years before the AGN shutting off and its feedback.  This
SFR is therefore an upper limit to the on going SFR.  Indeed, Davies
et al. (2007) found that the on going SFR in the nuclei of Markarian
231 and NGC1068 is probably very small, because of the small observed
Br$\gamma$ equivalent width within 0.1-0.5 kpc from the active
nucleus.

A correlation between $\dot E_{kin}$ and SFR would naturally emerge if
winds were supernova (SN) driven. The dashed line in
Fig. \ref{ekinsfr}, left panel, is the expectation for SN-driven
winds, by assuming 0.0066 SNe per solar mass of newly formed star
(Salpeter IMF), a total luminosity for each SN of $10^{51}$ ergs/s,
and a 10\% efficiency in releasing this luminosity into the ISM to
drive a shock.  The SN rate per solar mass is 0.0032 and 0.0083
$M_\odot^{-1}$ for a Scalo and Chabrier IMF, respectively (Somerville
\& Primack 1999, Dutton \& van der Bosh 2009). Therefore, SNe do not
seem powerful or numerous enough to drive most observed winds.

If the winds are AGN driven, a correlation between $\dot E_{kin}$ and
SFR would actually be expected because of the correlation between
$L_{bol}$ and SFR (Fig.  \ref{ekinsfr} right panel). Several authors
published correlations between AGN luminosity and SFR, whose scatter
is large, in particular at low AGN luminosities (e.g. Shao et
al. 2010, Rosario et al. 2012, Mullaney et al. 2012, Hickox et
al. 2014, Rodighiero et al. 2015).  As an example, Mullaney et
al. (2015) find that the distribution of the offset from the main
sequence $SFR/SFR_{MS}$ of X-ray selected AGN in CDFS follows a
log-normal distribution with $\sigma\sim0.6$ dex, nearly independent
on redshift. We plot in Fig. \ref{ekinsfr}, right panel, the expected
SFR based on the SFR$-L_{bol}$ correlations by Netzer (2009), Hickox
et al. (2014), z=0, and Hickox et al. (2014), z=2. It should be kept
in mind that these correlations concern the average SFR. It should
also be kept in mind that these correlations are probably driven by
scaling laws. Larger systems are more likely to have higher
luminosities, more powerful outflows and larger SFRs. What really
matters is the size of the system (also see Mancuso et al. 2016).

\begin{figure*}
\centering
\begin{tabular}{cc}
\includegraphics[width=8.5cm]{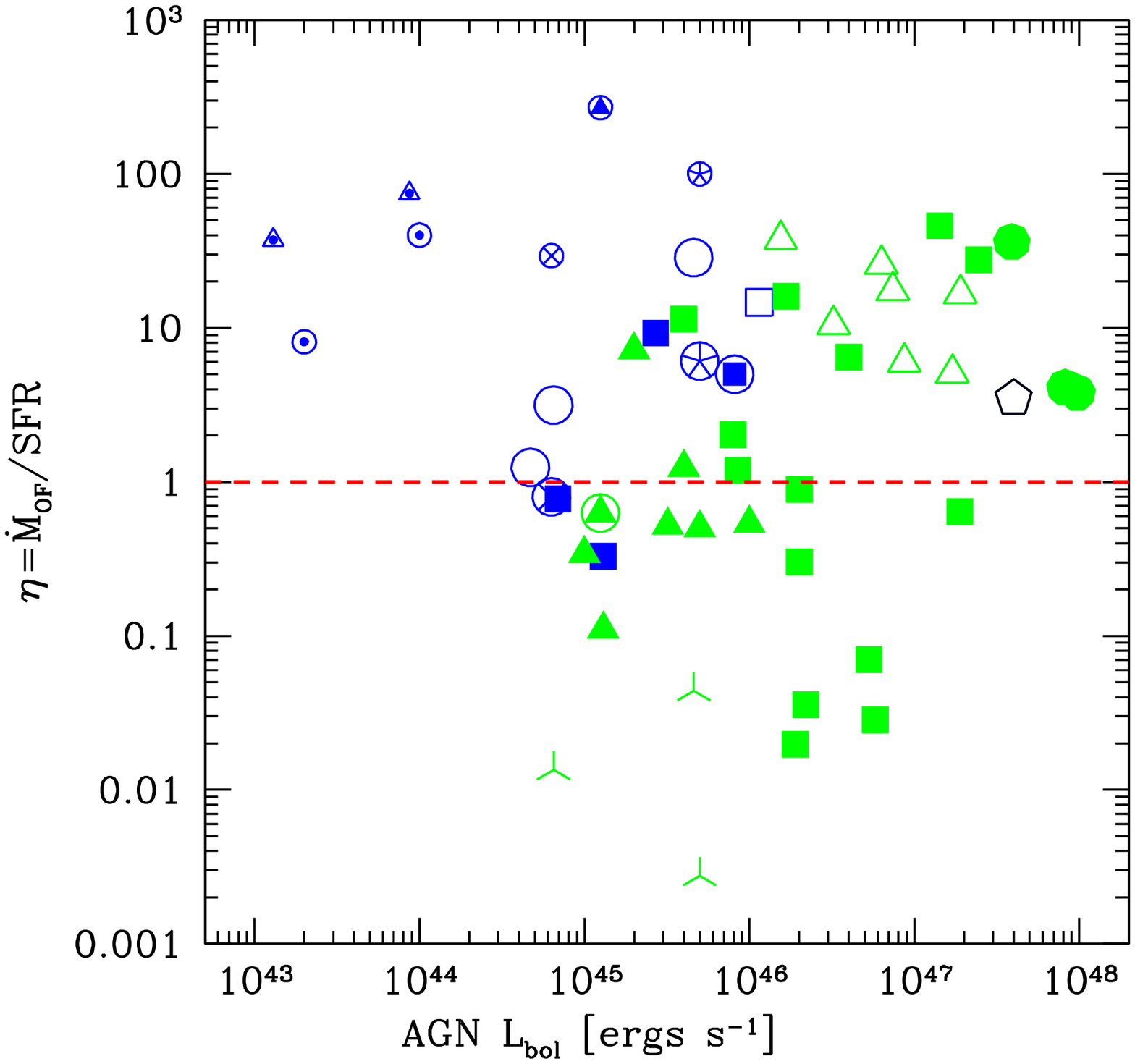}
\includegraphics[width=8.5cm]{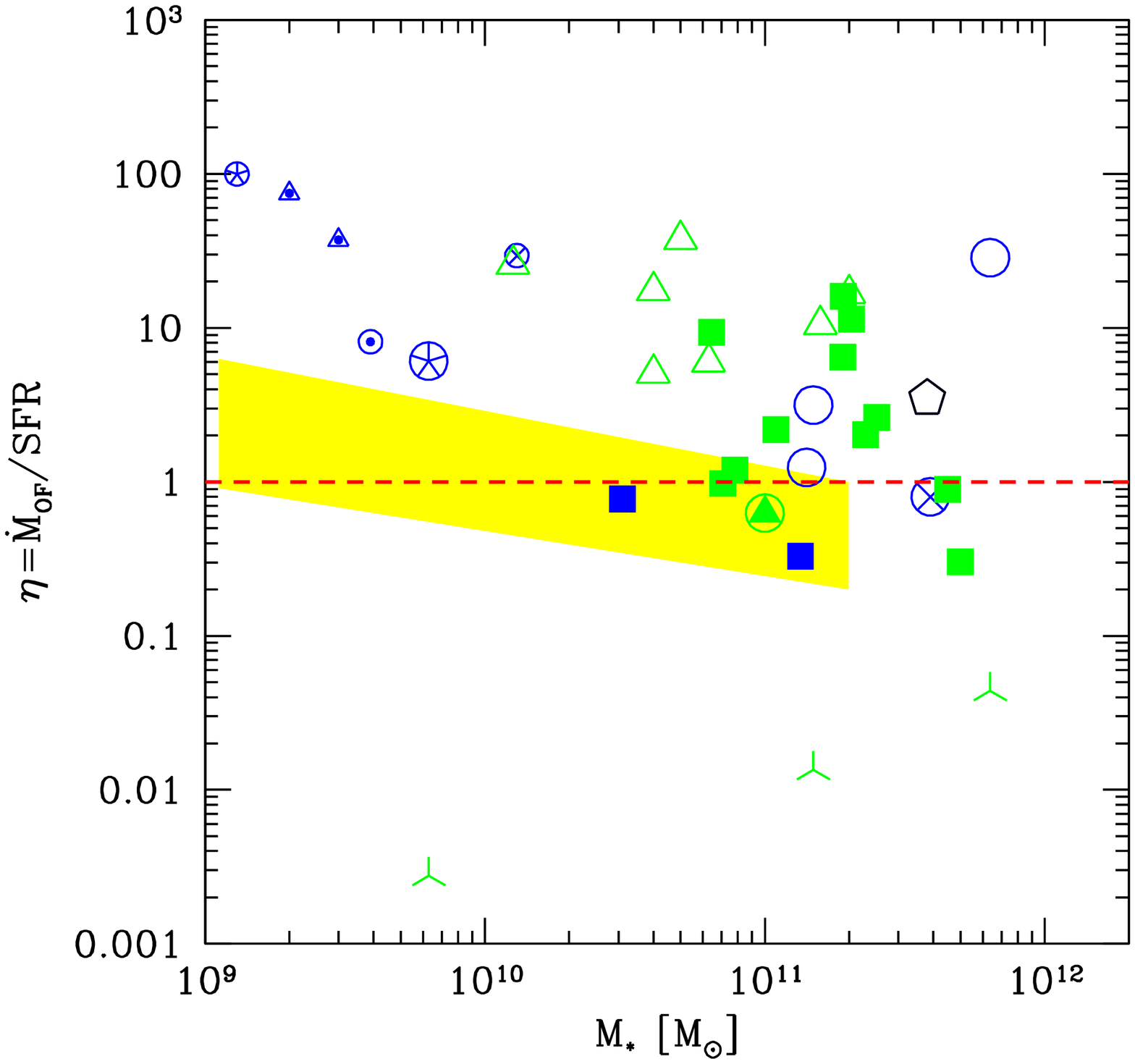}
\end{tabular}
\caption{The mass-loading factor $\eta=\dot M_{OF}$/SFR as a function
  of AGN bolometric luminosity [left panel], host galaxy stellar mass
  [right panel]. The yellow band in the right panel is the range found
  by Heckman et al. (2015)  for starburst driven galactic winds in a
  sample of local star-forming galaxies. Symbols as in Fig. \ref{lbolmdot}.}
\label{massload}
\end{figure*}

Figure \ref{massload} shows the mass-loading factor, $\eta=\dot
M_{OF}$/SFR, as a function of AGN bolometric luminosity and stellar
mass. The mass-loading factor of molecular winds is $>1$ in most
cases, and $>10$ in about half the cases. The median mass-loading
factor of ionised winds is $\approx 1$, with a rather large
distribution between 0.001 and 100. $\eta$ is not correlated with the
AGN bolometric luminosity while it is weakly anticorrelated with
stellar mass. The AGN driven wind mass-loading factors are
systematically larger than those of starburst driven winds in local
star-forming galaxies (Heckman et al. 2015, yellow band in
Fig. \ref{massload}).

\subsection{Molecular gas fractions and depletion timescales of AGN with massive winds}

For the sample of AGN with CO measurements we compute the
depletion timescale (i.e. the time needed to convert
all molecular gas into stars at the current star-formation rate),
$t_{dep}(SF)=M_{gas}/SFR$), and the molecular gas
fraction (i.e. the ratio of the molecular gas mass to stellar mass
$f_{gas}=M_{gas}/M_*$). We can then compare the distributions of
$t_{dep}$ and $f_{gas}$ to the corresponding Genzel et al. (2015)
scaling relations.  We use the equations in Whitaker et al. (2012) and
Genzel et al. (2015) to compute the specific SFR of the galaxy main
sequence (MS) as a function of redshift and stellar mass, $sSFR_{MS}$.

\begin{figure}
\includegraphics[width=8.5cm]{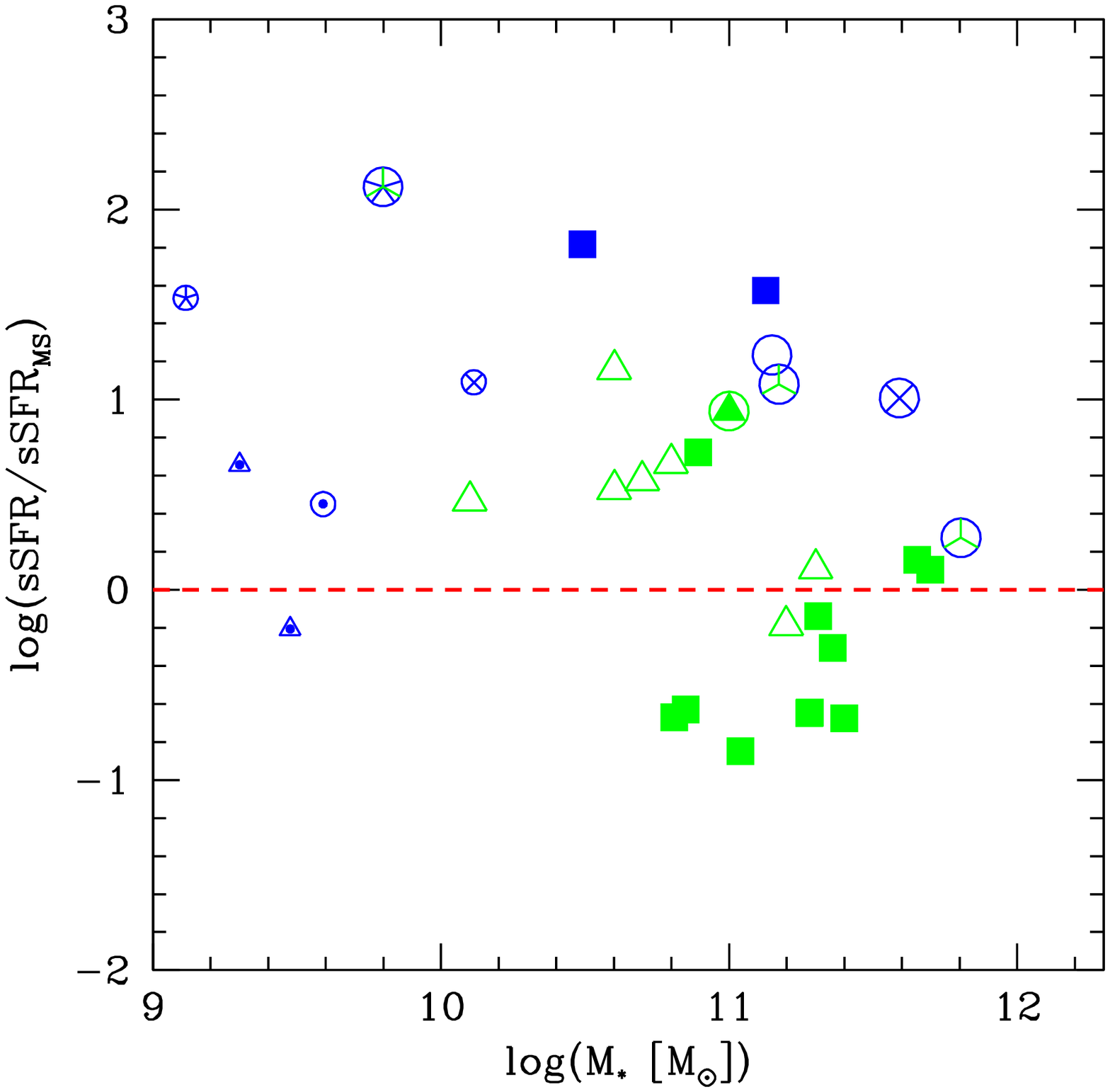}
\caption
{The offset from the galaxy main sequence (sSFR/sSFR$_{MS}$)
 as a function of the AGN host galaxy stellar mass. Symbols as in Fig. \ref{lbolmdot}.}
\label{distms}
\end{figure}

Fig. \ref{distms} shows the offset from the MS, $log(sSFR/sSFR_{MS})$,
as a function of the stellar mass for the samples of ionised and
molecular winds. This distribution is the result of assembling an
heterogeneous sample, with different selection criteria, and shows how
much the present sample is biased toward starbursts systems. In fact,
the first molecular outflows were found in local starburst galaxies
hosting an AGN (Fischer et al. 2010, Feruglio et al. 2010, Sturm et
al. 2011). Conversely, ionised outflows at z$\sim2$ are from samples
of SMGs or QSOs.  It is important to consider these different
selection criteria and distributions of source samples with respect to
the MS for the following discussion.

\begin{figure*}
\centering
\begin{tabular}{cc}
\includegraphics[width=8.5cm]{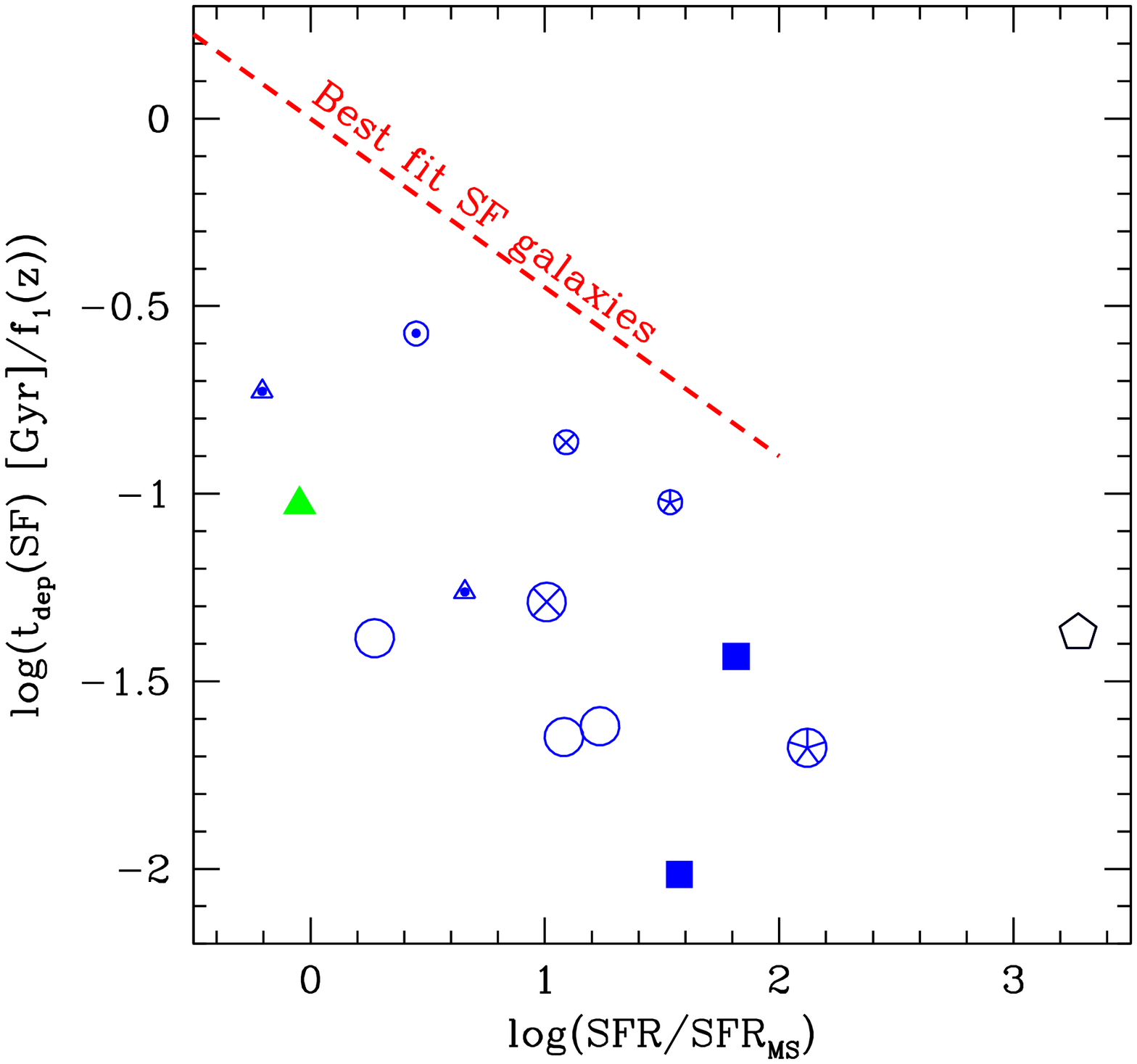}
\includegraphics[width=8.5cm]{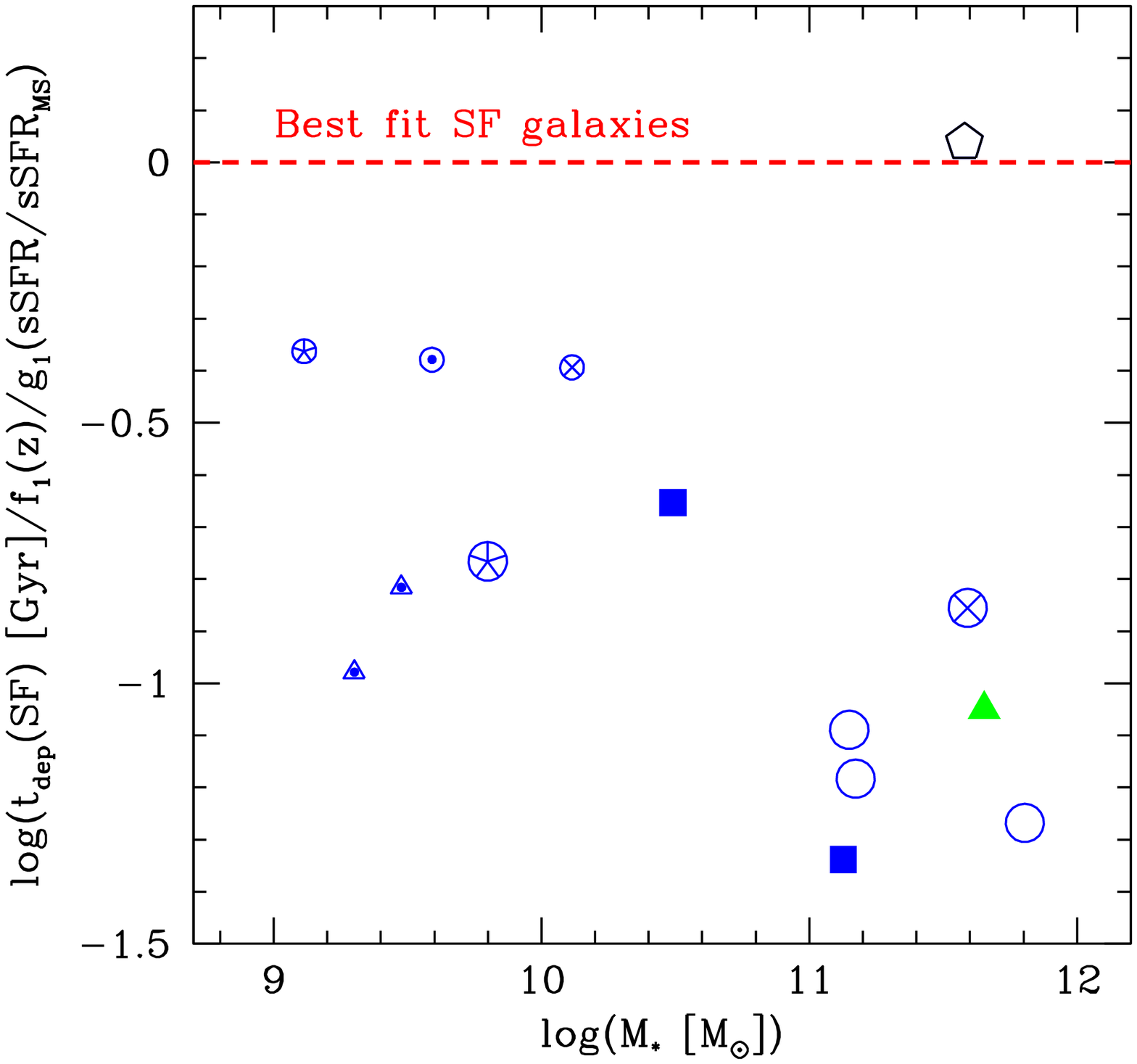}
\end{tabular}
\caption{[Left panel] The depletion timescale,
  $t_{dep}(SF)=M_{gas}/SFR$, as a function of the offset from the main
  sequence, after normalization to the mid-line of main-sequence at
  each redshift, by removing the redshift dependences with the fitting
  functions $f_1(z)$ in Table 3 of Genzel et al. (2015) corresponding
  to CO data, global distribution. [Right panel] The depletion
  timescale as a function of the galaxy stellar mass, after
  normalization to the mid-line of main-sequence, by removing the
  specific star-formation rate dependence with the fitting function
  g1($sSFR/sSFR(ms,z,M_*)$ in Table 3 of Genzel et al. (2015) for CO
  data, global distribution. Symbols as in Fig. \ref{lbolmdot}. The
  black pentagon marks J1148+5251 at z=6.4. The green triangle marks
  the QSO XID2028 with a detected ionised wind and measured molecular
  gas mass (Cresci et al. 2015, Brusa et al. 2015b). The red, dashed
  lines are the best linear fits to the log–log distributions of 500
  CO star-forming galaxies in Genzel et al. (2015).}
\label{tdep}
\end{figure*}

\begin{figure*}
\centering
\begin{tabular}{cc}
\includegraphics[width=8.5cm]{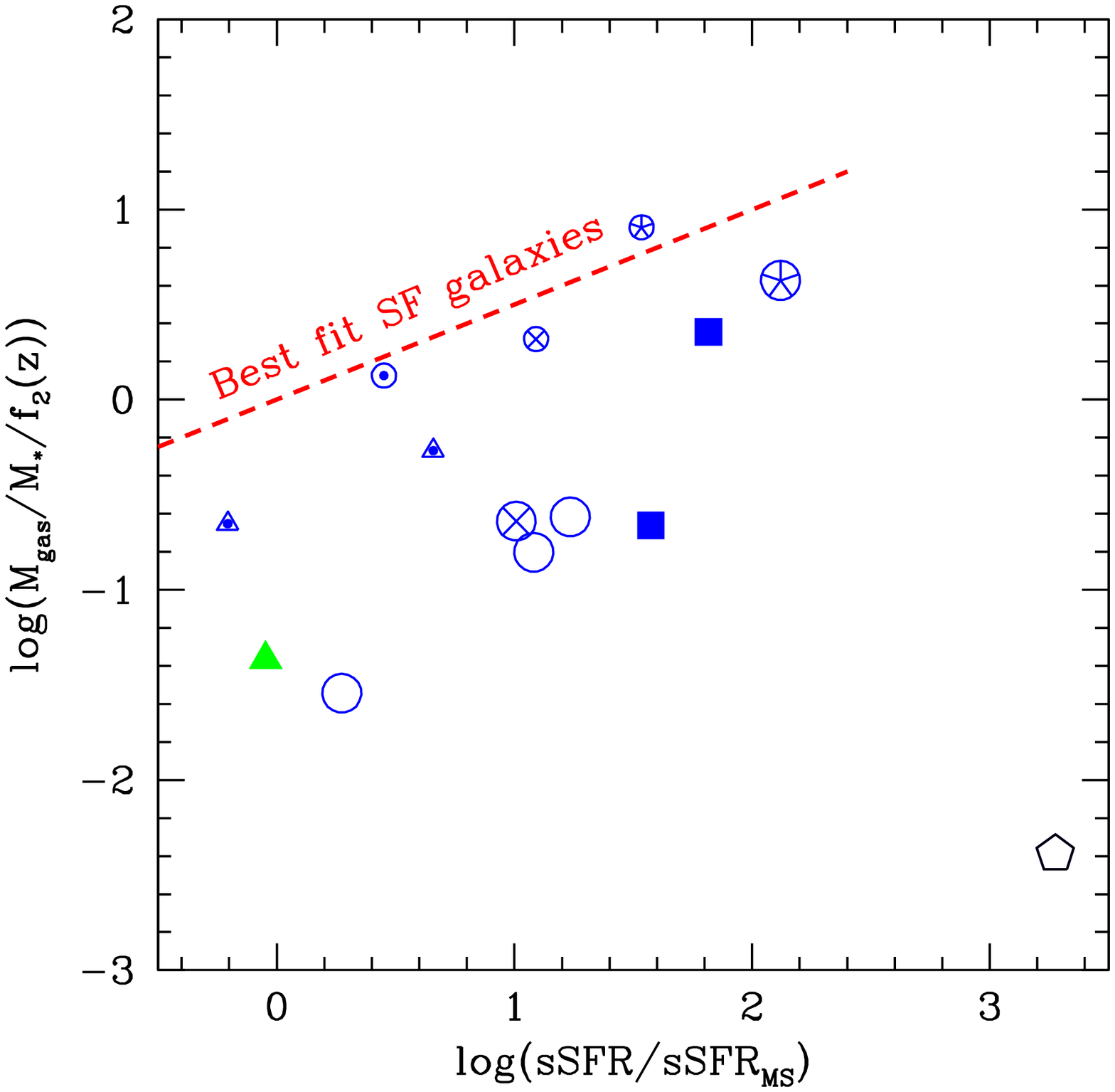}
\includegraphics[width=8.5cm]{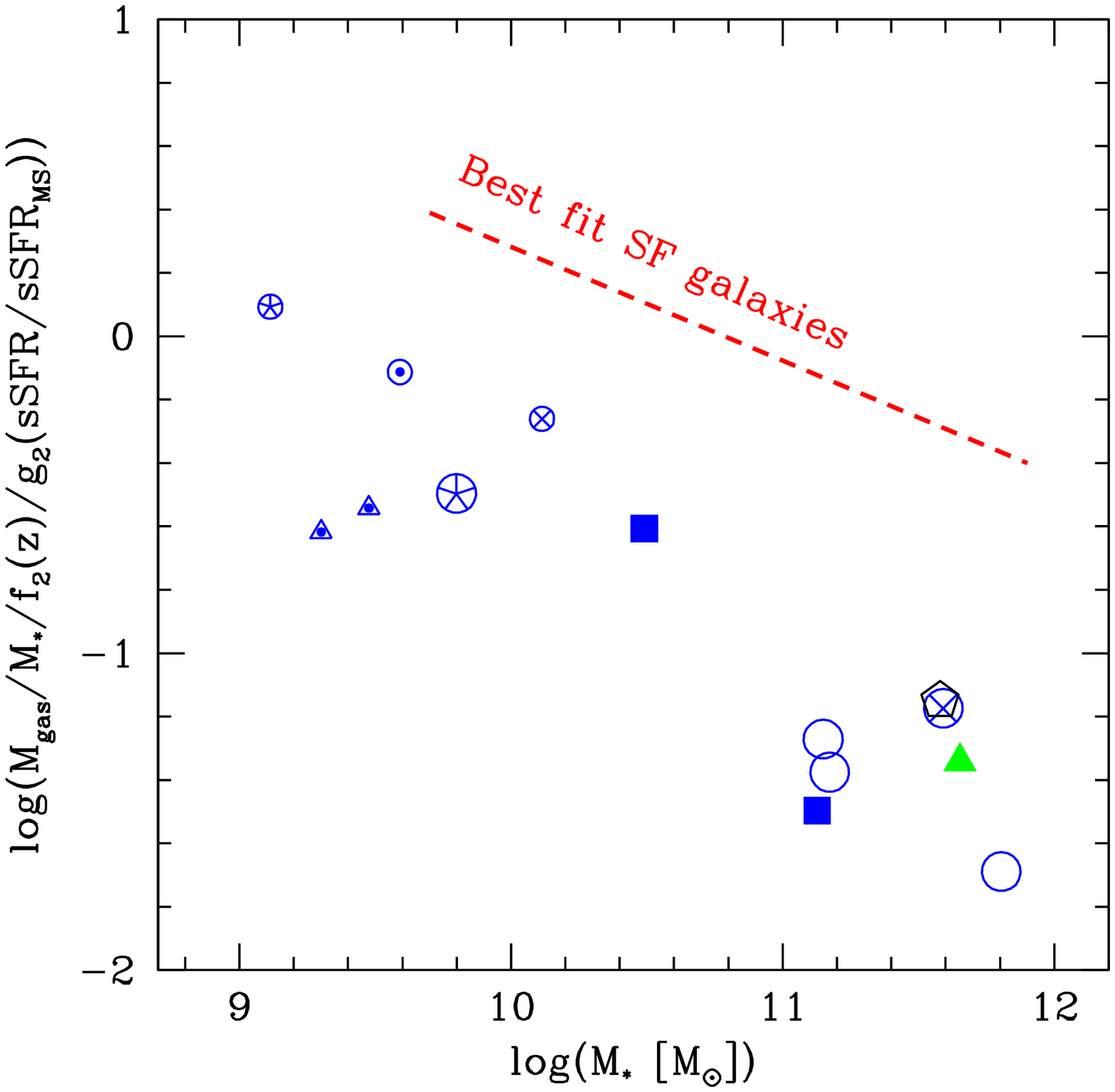}
\end{tabular}
\caption{[Left panel] The molecular gas fraction $f_{gas}=M_{gas}/M_*$
  as a function of the offset from the main sequence, after
  normalization to the mid-line of main-sequence at each redshift, by
  removing the redshift dependences with the fitting functions
  $f_2(z)$ in Table 4 of Genzel et al. (2015) corresponding to CO
  data, global distribution. [Right panel] The molecular gas fraction
  as a function of the galaxy stellar mass, after normalization to the
  mid-line of main-sequence, by removing the specific star-formation
  rate dependence with the fitting function g2($sSFR/sSFR(ms,z,M_*)$
  in Table 4 of Genzel et al. (2015) for CO data, global
  distribution. The black pentagon marks J1148+5251 at z=6.4.  The
  green triangle marks the QSO XID2028 with a detected ionised outflow
  and measured molecular gas mass (Cresci et al. 2015, Brusa et
  al. 2015b). The red, dashed lines mark the average correlations found
  by Genzel et al. 2015.}
\label{fgas}
\end{figure*}

Following Genzel et al. (2015), Figures \ref {tdep} and \ref{fgas}
show the depletion timescale and the molecular gas fraction as a function of the
the offset from the MS and stellar mass, after normalisation for the
trends with redshifts and offset from the MS (i.e the functions
$f_1(z)$ and $g_1(sSFR/sSFR_{MS})$ and $f_2(z)$ and
$g_2(sSFR/sSFR_{MS})$, Genzel et al. 2015). We find that the depletion
timescale, normalised for the offset from the MS and the trend with
the redshift is between -1 and -0.5 dex shorter than the average for
$M_*<10.5$, further reducing at higher stellar masses (the only point
close to average is the z=6.4 QSO J1148+5251, which has the most
uncertain estimate of both stellar mass and SFR).  The normalised gas
mass is also systematically smaller than the average found by Genzel
et al. (2015) (a factor of $\sim10$ at $M_*=10.5$). The normalised gas
mass reduces at high stellar masses with a slope similar to that of
the average ($\sim-0.5$, Genzel et al. 2015).

Brusa et al. (2015b) proposed that feedback due to strong winds in
massive AGN host galaxies may be the cause of the shorter depletion
time scales and smaller molecular gas measured in a z~1.5 obscured QSO
(XID2028, green triangle) and this may also be the case for the
galaxies in our sample.  However, we note that part of the offset of
galaxies hosting powerful molecular winds from average star-forming
galaxies may be due to the adopted conversion factor from L'(CO) to
$M_{gas}$ (we adopted $\alpha_{CO}=0.8$ for our sample, mostly made by
LIRGs and ULIRGs, while Genzel et al. (2015) use a complex conversion
function, that takes into account metallicity and density-temperature
dependence).  This may account for up to a factor of $\sim3-4$ in
$t_{dep}$ and $f_{gas}$. Even taking into account this correction,
most points in Figures \ref{tdep} and \ref{fgas} would fall short of
the Genzel et al. (2015) average values, in particular at high stellar
masses. It should also be considered that the Genzel et al. (2015)
averages themselves (i.e. the average depletion timescales and gas
fraction after the subtraction of the trends with redshift and offset
from the main sequence), may well be affected by uncertainties. For
example, the Sargent et al. (2014) parameterisation results in a
depletion timescale a factor of $\sim2$ shorter than the Genzel et
al. (2015) one for galaxies above the MS.

\begin{figure}[h!]
\includegraphics[width=8.5cm]{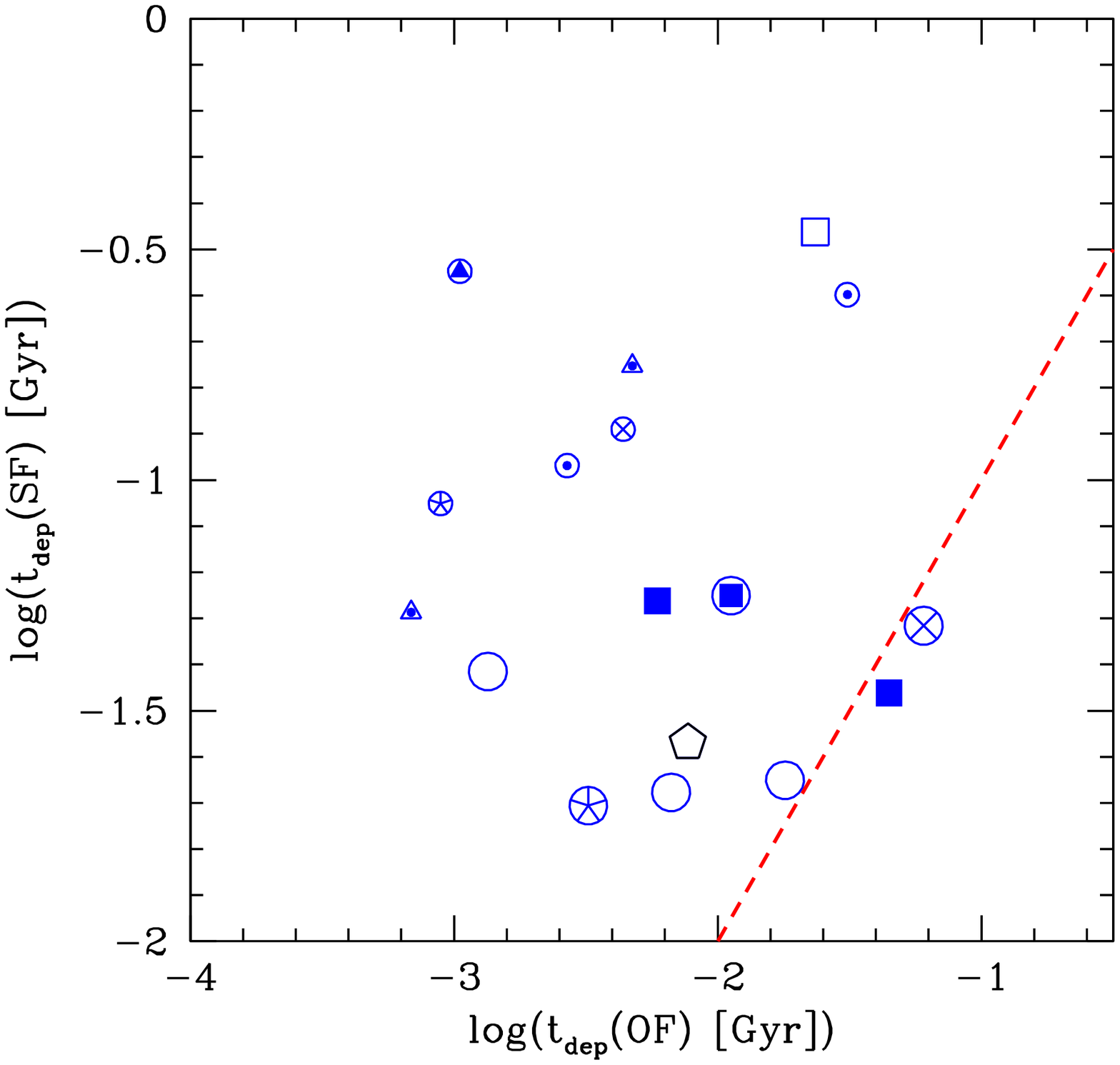}
\caption{The SFR depletion timescale, $t_{dep}(SF)=M_{gas}/SFR$, as a
  function of the wind depletion timescale, $t_{dep}(OF)=M_{gas}/\dot
  M_{OF}$. The red dashed line is the depletion time scale due to star
  formation at the measured rate.}
\label{tdep2}
\end{figure}

Figure \ref{tdep2} shows the star-formation against the outflow
depletion timescales, $t_{dep}(OF)=M_{gas}/\dot M_{OF}$. At face
value, in most systems, and in particular the six galaxy nuclei,
outflows are powerful enough to deplete the galaxy molecular gas
reservoir in a timescale shorter than that needed to exhaust it by
forming stars at the measured rate (red dashed line). This assumes
that the molecular winds are not blocked at some distance and do not
dissolve out.  Pressure-confined molecular clouds may, however,
dissolve out as the wind expands, and CO may be efficiently
photo-dissociated by the UV radiation, since self shielding will be
strongly reduced at low densities. The best studied molecular wind so
far (Markarian 231, Feruglio et al. 2015) has a size of $\sim1$
kpc. At this distance the mass in outflow strongly reduces, while its
velocity remains nearly constant, suggesting that a large part of the
molecular gas leaves the flow during its expansion. This molecular gas
may rain back onto the nucleus or the disk, replenishing the gas
reservoirs.

\section{AGN winds in a cosmological framework}

We now attempt to put the results from the previous section in the
cosmological evolution framework. This will enable us to assess the
relative importance of AGN driven winds on the average cosmological
star-formation, accounting for the fraction of galaxies which are
caught in the AGN phase. This fraction can be as low as 1\% in the
local Universe, and up to 30\% at z=2 (Brusa et al. 2009, Fiore et
al. 2012, Bongiorno et al. 2012).  We first summarise the results on
the evolution of AGN and galaxies luminosity densities, we then link
SMBH accretion to star-formation, and finally estimate the AGN wind
mass loading factor density as a function of the cosmic time.

\subsection{The evolution of the AGN luminosity density}
 
\begin{figure*}[t!]
\centering
\begin{tabular}{cc}
\includegraphics[width=8.5cm]{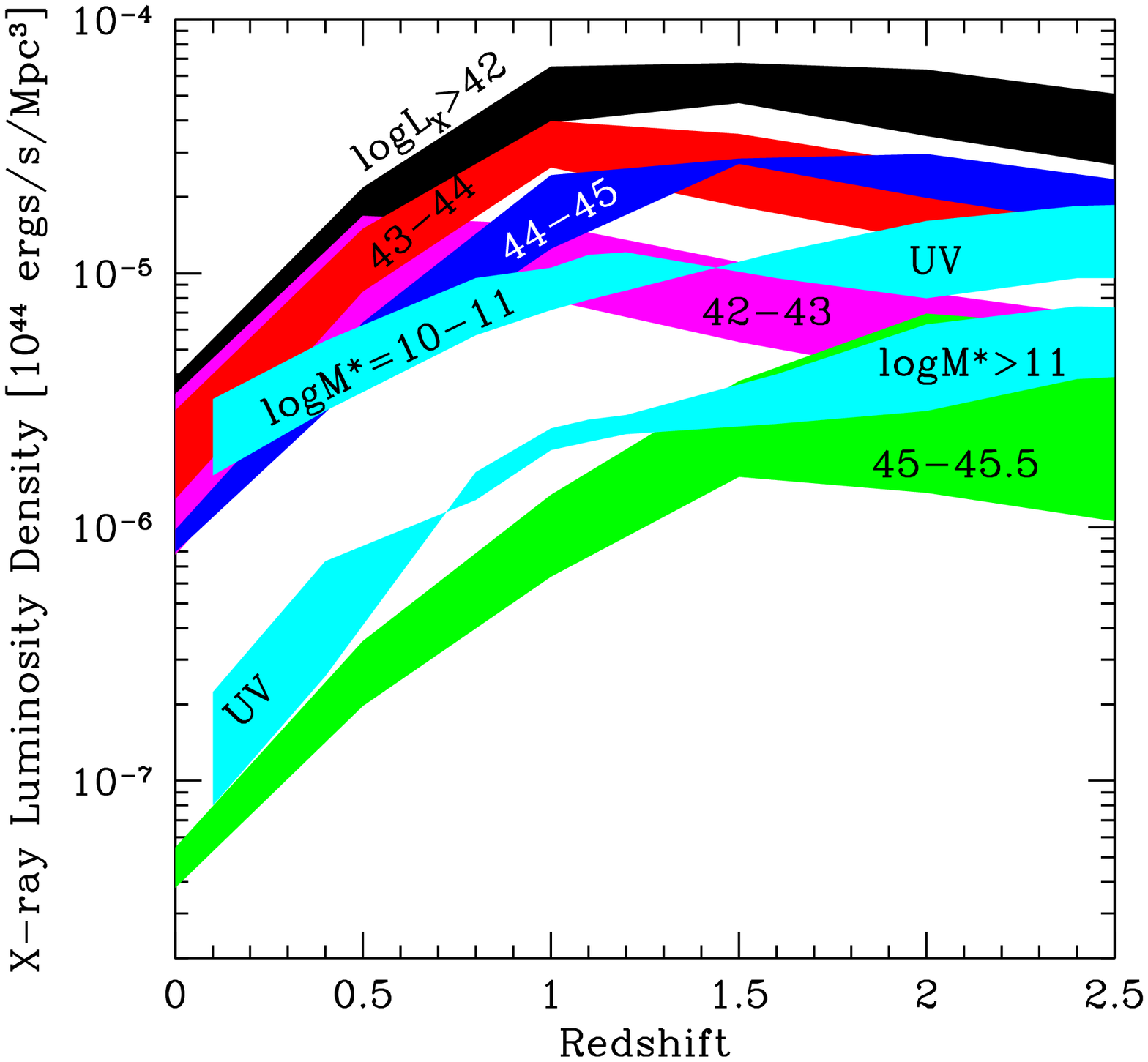}
\includegraphics[width=8.5cm]{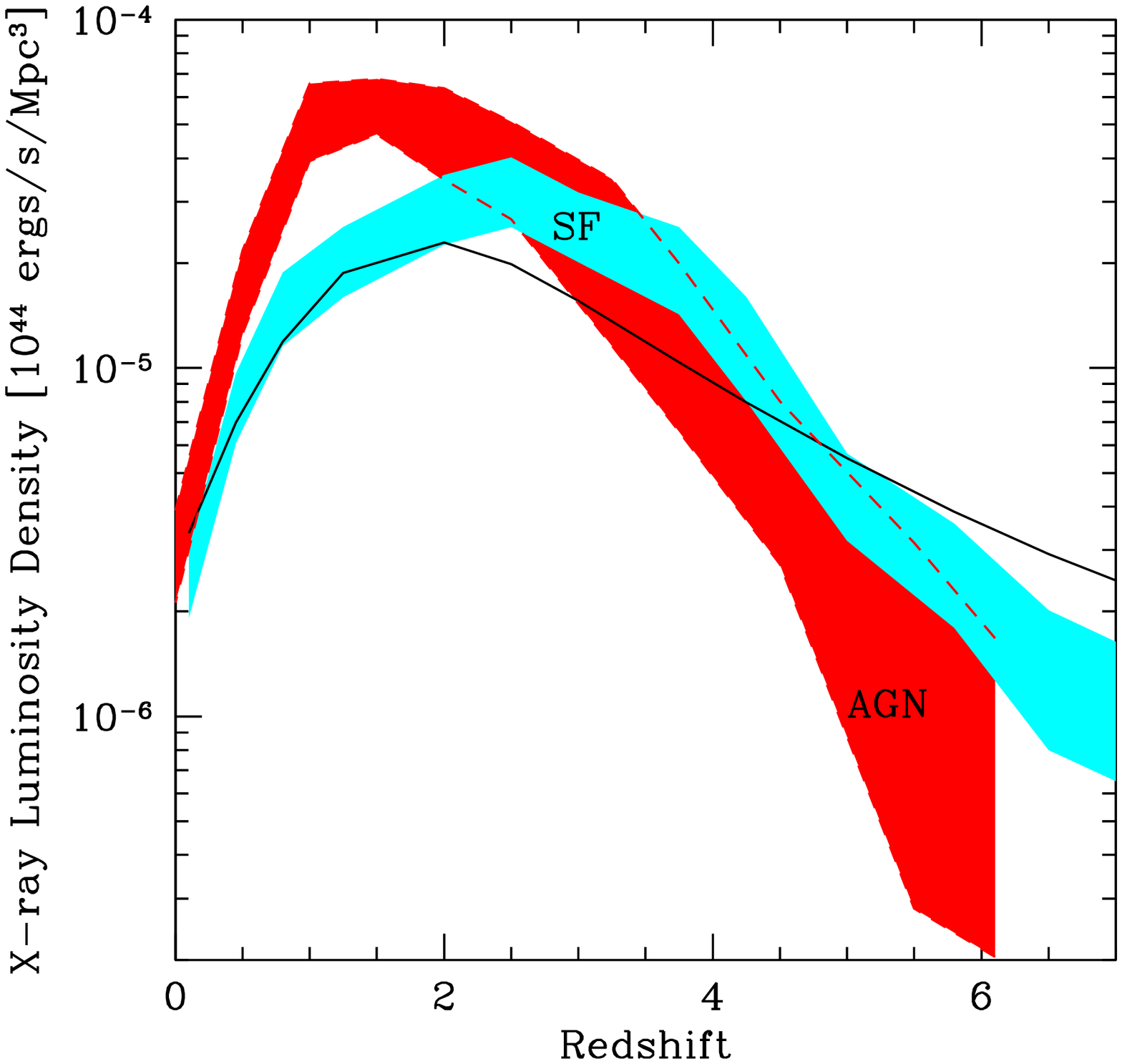}
\end{tabular}
\caption{The evolution of the X-ray 2-10 keV AGN luminosity density
  for different AGN luminosities.  [Left panel]: AGN luminosity
  density for AGN, split in ranges of 2-10 keV luminosity, as labeled.
  Cyan bands show the SF luminosity density as estimated by Santini et
  al. (2009) and Gruppioni et al. (2015), divided by a factor of 1000
  for plotting purpose, for galaxies with log$M_*$=10-11 and
  log$M_*>11$.  [Right panel:] total AGN luminosity density for AGN
  with L(2-10)=$>10^{43}$ ergs/s (red band). SF luminosity density as
  estimated by Santini et al. (2009), Gruppioni et al. (2015), and
  Bouwens et al. (2011, 2015) and divided by a factor 1000 for
  plotting purposes (cyan band). Average SF luminosity density as
  estimated by Madau \& Dickinson (2014) (black solid line). }
\label{dens}%
\end{figure*}

We plot in Fig. \ref{dens} the evolution of the X-ray 2-10 keV AGN
luminosity density for different AGN luminosities. AGN as faint as
L(2-10 keV)=$10^{42}$ ergs/s can be detected by Chandra in ultra-deep
surveys, even up to z=2.5-3. Brighter AGN with L(2-10)=$>10^{43}$
ergs/s are observable up to z=6. For this reason, we provide two
distinct plots, one including $10^{42}$ ergs/s AGN up to z=2.5 and
another including L(2-10)=$>10^{43}$ ergs/s up to z=6.  The shaded areas
account for the uncertainties.  AGN samples at z$<$3 include today
several thousands objects, resulting into small statistical errors on
the luminosity functions up to this redshift. In particular, the
statistical error is smaller than the systematic error due to the
different assumptions that authors make to account for selection
effects. At z$<3$ the shaded areas bracket the determinations of La
Franca et al. (2005), Ebrero et al. (2009), Aird et al. (2010), and
Ueda et al. (2014).  Conversely, X-ray selected AGN samples at z $>3$
are still relatively small, including hundred objects at z=3-4, a few
dozen objects at z$>$4, and a few at z$>6$. Optically selected AGN at
z$>$6 are relatively rare too, with only $\sim$a few dozen luminous
QSOs at z$>$6 known so far.  As a consequence, the main error source
on the AGN luminosity functions at the low-medium luminosities sampled
by X-ray surveys at z$>3$, and by optical surveys at z$>6$, is the
statistical error. At z$>3$ we used the Fiore et al. (2012), Ueda et
al. (2014), Georgakakis et al. (2015), Aird et al. (2015), Kalfountzou
et al. (2014), Vito et al. (2014), Marchesi et al. (2016), and
Puccetti et al. (in preparation) AGN luminosity functions. The shaded
areas account for both statistical and systematic errors. The left
panel of Fig. \ref{dens} clearly shows the downsizing of AGN X-ray
luminosity density, with AGN of X-ray luminosity $10^{43}-10^{44}$
ergs/s peaking at z=1, and AGN of X-ray luminosity $>10^{45}$ ergs/s
peaking at z=$\gs2$. The total AGN X-ray luminosity density peaks at
z=1-2 (right panel). The right panel of Figure \ref{dens} also shows
the galaxy UV luminosity density, scaled by a factor $10^3$ (from
Bouwens et al. 2011, 2015, Santini et al. 2009, Gruppioni et al. 2015,
and Madau \& Dickinson 2014).  Note that the total galaxy UV
luminosity density peaks at z=2-3, a redshift higher than that of the
peak of the AGN X-ray luminosity density.

\begin{figure}[h!]
\centering
\includegraphics[width=8.5cm]{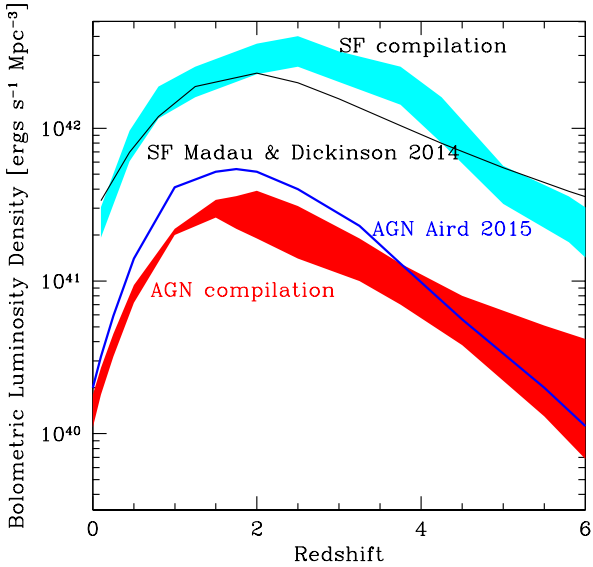}
\caption{The evolution of the AGN bolometric luminosity density: the
  red band has been computed from a compilation of X-ray luminosity
  functions integrated in the range log$L_X=42-45$, se text for
  details, and assuming the Marconi et al. (2004) bolometric
  correction; the blue solid line is the Aird et al. (2015)
  determination. Fort the UV luminosity density the cyan band is the
  average of a compilation from Santini et al. (2009), Gruppioni et
  al. (2015), Bouwens et al. (2011, 2015). The black solid line is the
  Madau \& Dickinson (2014) determination.}
\label{denslbolz}%
\end{figure}

While the galaxy UV luminosity is linearly correlated with the SFR,
the X-ray luminosity is not a good proxy of the AGN bolometric
luminosity, and thus of the SMBH gas accretion rate. The relationship
between X-ray luminosity and the AGN bolometric luminosity is complex
and may depend on several parameters. Several authors suggested a
polynomial scaling between log$L_X$ and log$L_{bol}$ (Marconi et
al. 2004, Hopkins et al. 2006). Others suggest a scaling with the
Eddington ratio (Vasudevan \& Fabian 2007, Jin et al. 2012). As an
example, we adopt the Marconi et al. (2004) scaling to calculate the
evolution of the AGN bolometric luminosity density given in Table 2
and plotted in Fig. \ref{denslbolz} (we integrate the X-ray AGN
luminosity functions in the luminosity range $10^{42}-10^{45}$ ergs/s). This
is compared with the similar determination of Aird et al. (2015) and
with the evolution of the UV luminosity density. Our determination of
the AGN bolometric luminosity density based on a compilation from the
papers quoted above falls short by up to a factor 30\% from the Aird
et al. (2015) determination (total AGN luminosity density, included
the contribution of Compton thick AGN) at z $\ls3$. Above this
redshift it is consistent with the Aird et al. (2015) determination
within the, rather large, uncertainties. The AGN bolometric luminosity
density is $\sim10$ times smaller than the UV luminosity density at
all redshift. The shape of the AGN bolometric and UV luminosity
density are similar, both peaking at z=1.5-2.5. Differences are
smaller than the systematic differences between different
determinations of AGN bolometric (compilation in this work vs. Aird et
al. 2015) and UV (compilation in this work vs. Madau \& Dickinson
2014). We remark here that these results are obtained using the
Marconi et al. (2004) bolometric correction to convert X-ray to
bolometric luminosity, while Aird et al. (2015) used the correction
provided by Hopkins et al. (2006).  Adopting other scalings, for
example assuming a more complex relationship between bolometric
luminosity, SMBH mass and Eddington ratio, would produce somewhat
different results.  We further investigate this issue in the next
sections.

\begin{table}
\begin{minipage}[!h]{1\linewidth}
\centering
\caption{AGN bolometric luminosity density evolution}
\begin{tabular}{lcc} 
\hline
Redshift & logL(min) & logL(max) \\
\hline
         0.1    &     40.1  &    40.4 \\
        0.25    &     40.6  &    40.7 \\
         0.5    &     40.8  &    41.0 \\
           1    &     41.0  &    41.3 \\
         1.5    &     41.2  &    41.5 \\
           2    &     41.3  &    41.6 \\
         2.5    &     41.4  &    41.5 \\
        3.25    &     41.3  &    41.3 \\
        3.75    &     40.9  &    41.1 \\
         4.5    &     40.5  &    40.9 \\
         5.5    &     40.2  &    40.7 \\
         6.0    &     40.0  &    40.6 \\
\hline
\hline
\end{tabular}
\end{minipage}
\end{table}

\subsection{SMBH accretion and star-formation}

A complementary approach to compare SMBH accretion and star-formation
is to self-consistently evolve the SMBH mass function via the
continuity equation (Cavaliere et al.\ 1971; Small \& Blandford 1992):

\begin{equation}
\frac{\partial n_{\rm BH}}{\partial
t}(M_{\rm BH},t)=-\frac{\partial (\langle \dot{M}_{\rm BH}\rangle
n_{\rm BH}(M_{\rm BH},t))}{\partial M_{\rm BH}}\, 
    \label{eq|conteq}
\end{equation}

$n_{\rm BH}(M_{\rm BH},t)$ is the number of SMBHs of mass \mbh\ at
time $t$ and $\langle \dot{M}_{\rm BH}(M_{\rm BH},t) \rangle$ is the
\emph{mean} accretion rate, averaged over the active and inactive
populations, of all SMBHs of mass \mbh\ at time $t$.  While
Eq.~(\ref{eq|conteq}) neglects any contribution from SMBH mergers, the
latter process does not impact the mean accretion rate but it mainly
alters the redistribution of the mass function (Shankar et al. 2009).
The average growth rate of all SMBHs can be computed by convolving the
probability to radiate at a given fraction $\lambda=L/L_{\rm Edd}$ of
the Eddington luminosity $P(\lambda|M_{\rm BH},z)$, and the overall
probability, or ``duty cycle'', to be active $U(M_{\rm BH},z)$

\begin{equation}
\langle \dot{M}_{\rm BH}\rangle=\int d\log \lambda \, P(\lambda|M_{\rm
BH},z)\lambda \, U(M_{\rm BH},z)\,  \frac{M_{\rm BH}}{t_{\rm s}}\, ,
\label{eq|MdotAve}
\end{equation}

\noindent
where $t_s$ is the AGN Salpeter timescale and the integral extends
over all allowed values of $\lambda$.  The input Eddington ratio
distributions are motivated by a variety of independent observational
probes, while the duty cycle is self-consistently re-computed at each
time $t$ from the ratio between the AGN luminosity function and SMBH
mass function at the previous time step (the full methodology and
numerical details can be found in, e.g., Shankar et al. 2013, and
references therein).

\begin{figure*}
\centering
\begin{tabular}{cc}
\includegraphics[width=8.5cm]{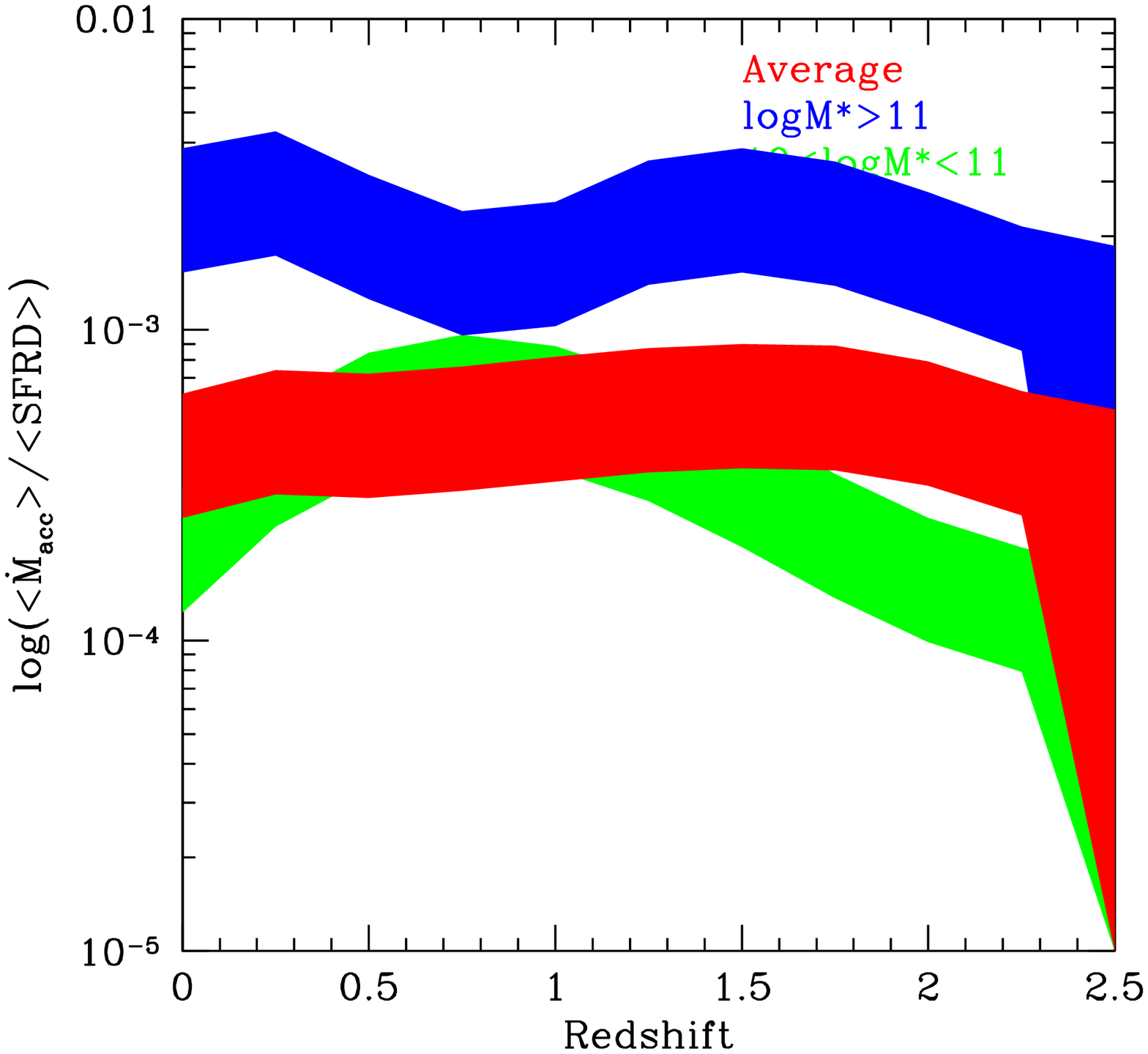}
\includegraphics[width=8.5cm]{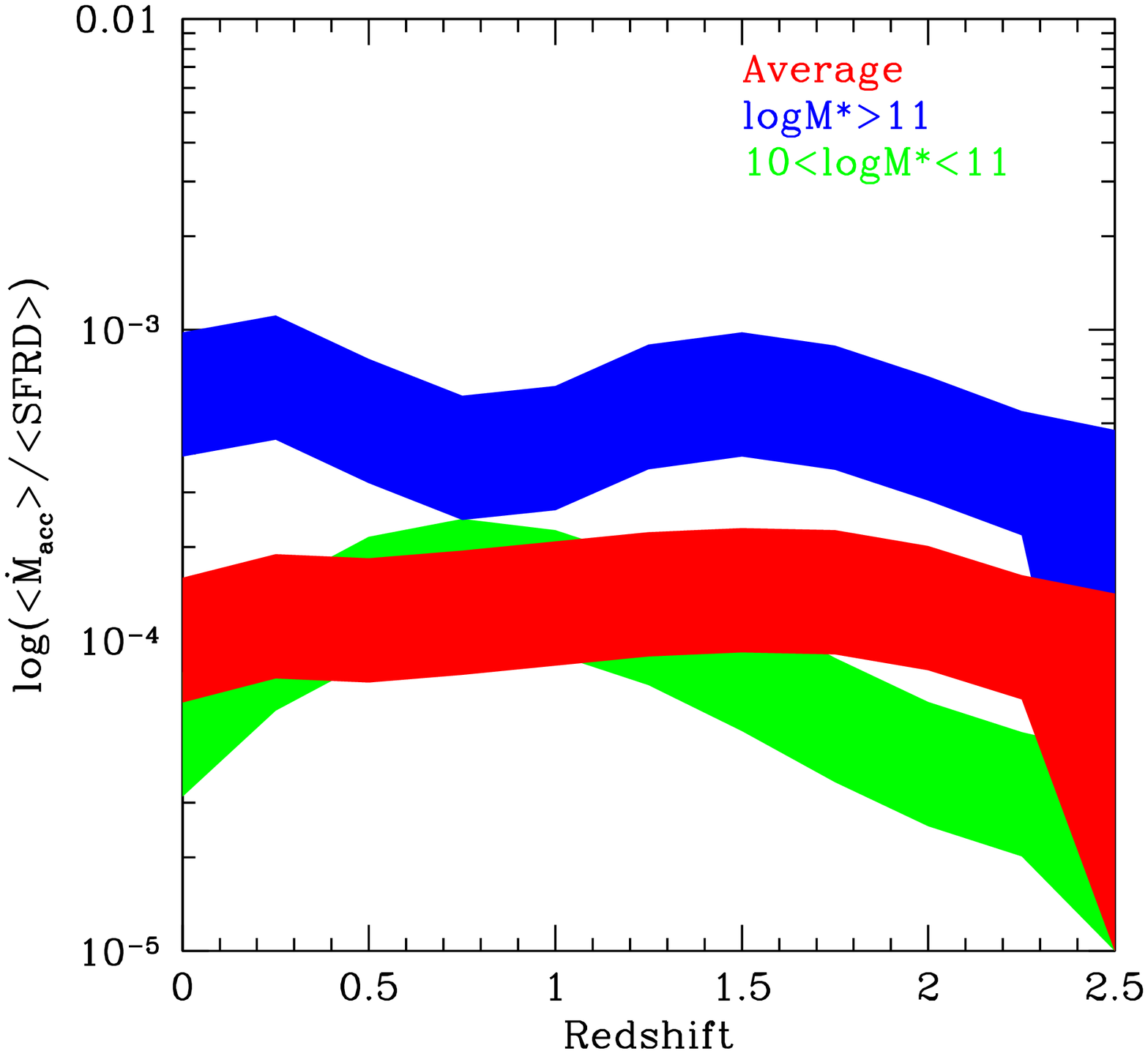}
\end{tabular}
\begin{tabular}{cc}
\includegraphics[width=8.5cm]{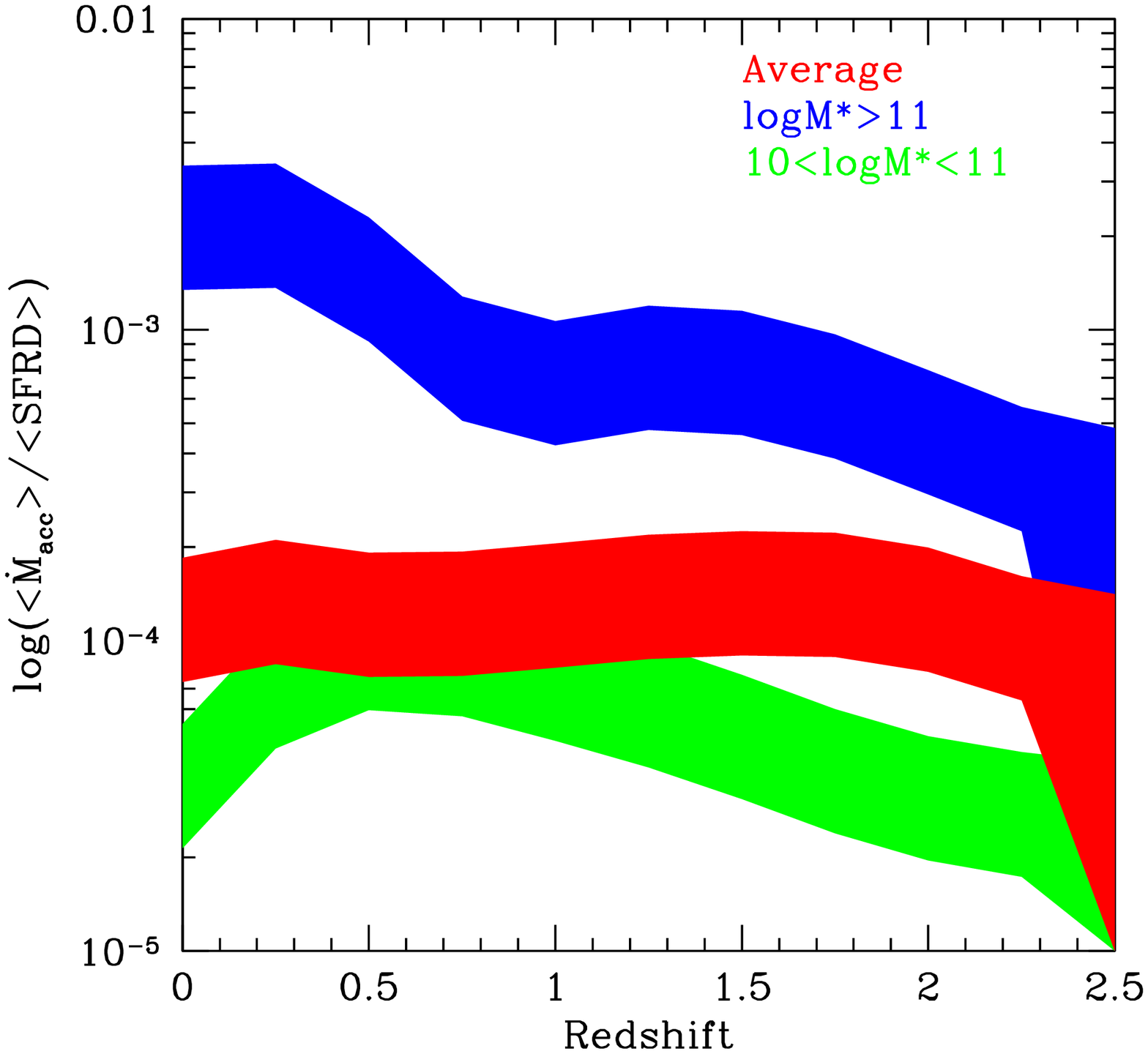}
\includegraphics[width=8.5cm]{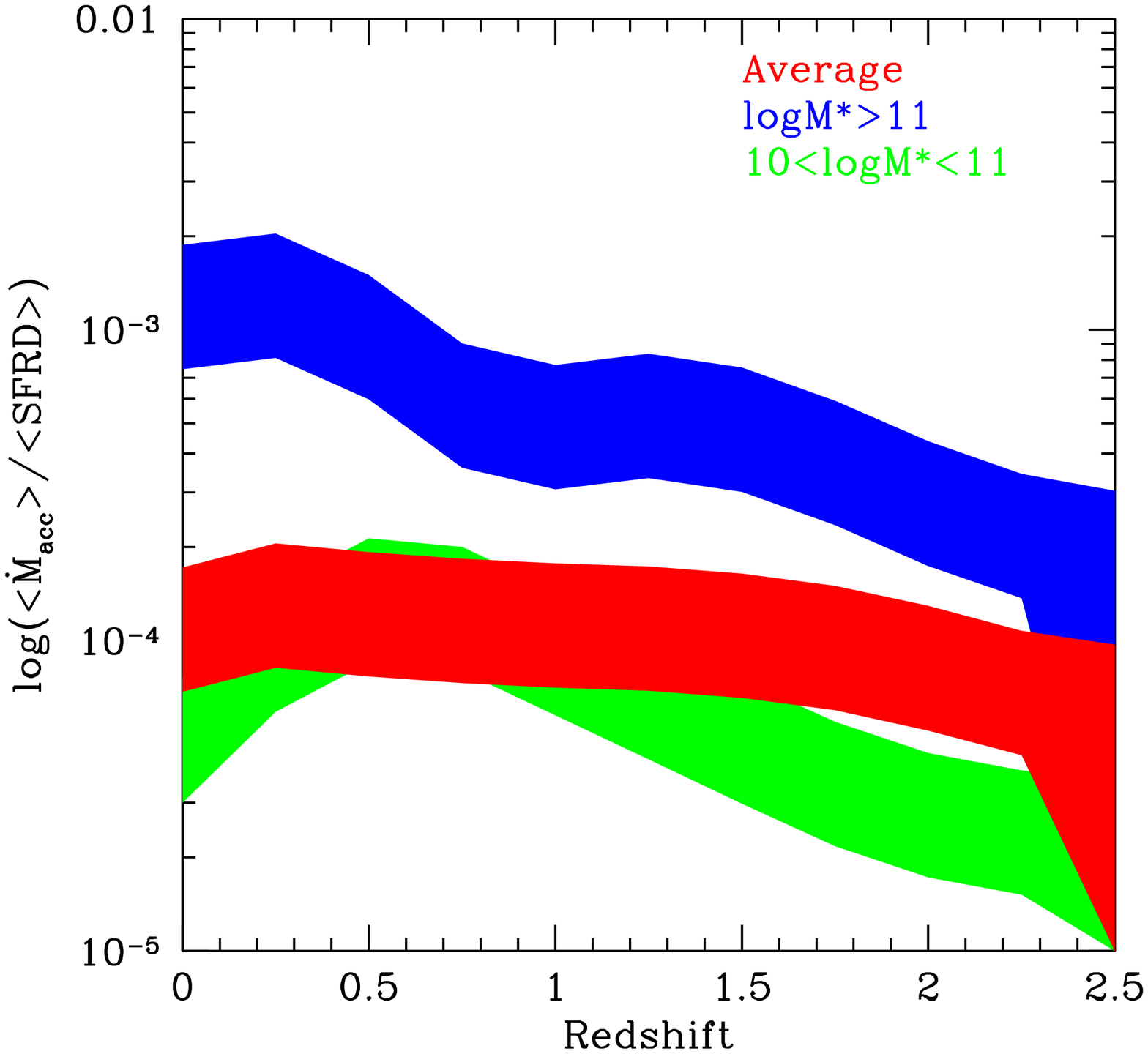}
\end{tabular}
\caption{The ratio between the SMBH accretion rate and SFR densities
  as a function of the redshift.  [Upper-left panel]: constant
  radiative efficiency $\epsilon=0.06$ and constant Eddington ratio
  distribution at all redshifts. [Upper-right panel]: constant
  radiative efficiency $\epsilon=0.2$ and constant Eddington ratio
  distribution at all redshifts. [Lower-left panel]: an evolving
  Eddington ratio (eg. 15 in Shankar et al. 2013),
  $\epsilon=0.2$. [Lower-right panel]: mass-dependent radiative
  efficiency ($\epsilon$ from 0.05 to 0.4 linearly with BH black hole
  mass at all redshifts and evolving Eddington ratio.}
\label{maccsfr}%
\end{figure*}

It has been already emphasized in the literature that the average SMBH
accretion rate density has a redshift dependence morphologically
similar to the cosmological SFR density (Marconi et al. 2004; Merloni
et al. 2004; Silverman et al. 2008; Zheng et al. 2009; Shankar et
al. 2009).  We provide in Fig.~\ref{maccsfr} our estimate of the ratio
between SMBH growth and SFR density. We first extract \mdotbh\ from
the continuity equation models of Shankar et al. (2013), and bin it in
the ranges $7<\log \mbhe /\msune <8$, $8 <\log \mbhe /\msune <9$, and
$\log \mbhe /\msune >9$, which implies integrating it over the
appropriate mass range of active SMBH mass function at all epochs. We
thus convert an accretion rate, measured in $\msune{\, yr^{-1}}$ to a
SMBH accretion rate density, measured in $\msune{\, yr^{-1}\,
  Mpc^{-3}}$. We then divide the accretion rate density for the
average SFR density.  Finally, we take the SFR densities in the
stellar mass ranges $10 <\log \mstare/\msune<11$ and $\log
\mstare/\msune>11$ and relate then to the accretion rate densities
from SMBH of masses $7<\log \mbhe /\msune <8$ and $\log \mbhe /\msune
>8$, respectively.  This allows us to infer an approximate
\emph{mass-dependent} correlation between \mdotbh\ and SFR density.
All models predict, on average, a nearly constant ratio in time of the
SMBH mean accretion and SFR (red shaded area).  The top panels
corresponds to constant Eddington ratio distribution at all redshifts
and constant radiative efficiency $\epsilon=0.06$ (left) and
$\epsilon=0.2$ (right). Note that the average ratio between the SMBH
accretion rate density and SFR density is about $3\times10^{-4}$ and
$10^{-4}$ for $\epsilon=0.06$ and $\epsilon=0.2$, respectively. Note
that the latter ratio is consistent with the {\it intrinsic} value
computed by Shankar et al. (2016) for the same $\epsilon$. At all
redshifts the ratio is higher for massive galaxies. The peak ratio of
less massive galaxies is at z=0.5-1 (which is also the redshift at
which the density of low luminosity AGN (L$_{2-10keV}\ls 10^{44}$
ergs/s) peak). Similar conclusions apply to the other models analyzed:
evolving Eddington ratio (lower-left panel), and mass-dependent
radiative efficiency (lower-right panel).  Interestingly, the strong,
apparent co-evolution between accreting SMBHs and galaxies appears to
break down for the least massive galaxies, while the most massive
galaxies tend to align with a ratio of $5-10\times10^{-4}$ or higher.

\subsection{The evolution of the AGN wind mass-loading factor}

The remarkable correlation between the AGN bolometric luminosity,
$L_{bol}$, AGN wind mass outflow rate, $\dot M_{OF}$, and kinetic
power, $\dot E_{kin}$ (see Fig. \ref{lbolmdot} and Section 2)
suggests that the AGN bolometric luminosity density can be converted
to a density of wind mass outflow rate and kinetic power.  This can
then be divided by the SFR density to compute an ``average''
mass-loading factor as a function of the redshift (under the
assumption that:

\begin{equation}
<\eta>=<\dot M_{OF}D/SFRD>\sim<\dot M_{OF}D>/<SFRD>).  
\end{equation}

To this purpose, we converted the AGN bolometric luminosity density
into a density of the AGN mass outflow rate using Monte Carlo
realisations.  More in detail, we first randomly chose a bolometric
luminosity following the luminosity function distribution in each
given redshift bin, and then convert it into mass outflow rate
assuming $\dot M_{OF} \propto L_{bol}^{0.76}$ (baseline scaling), and
normalization consistent with the findings for molecular winds in
Section 2 (we used the scaling $log\dot M_{OF}=0.76*logL_{bol}-32$,
dashed line Fig. 1 left panel). We remark that this scaling refers to
a {\it biased} sample of local AGN, and we assume that the same
scaling holds at all redshifts.  To study how much our conclusions
depend on the exact form of the scaling, we calculated the mass
outflow rate densities also adopting two different scalings: a square
root scaling between the AGN mass outflow rate and the bolometric
luminosity and a linear scaling. We first discuss the results obtained
with the baseline scaling, and then comment on the differences with
respect to the flatter and steeper scalings.

\begin{figure}
\centering \includegraphics[width=8.5cm]{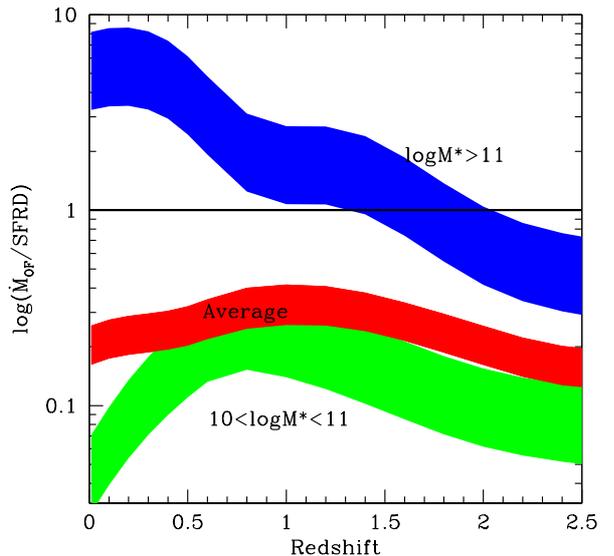}
\caption{The average mass-loading factor $<\eta>=<\dot M_{OF}D>/<SFRD>$ as a
  function of the redshift. Red=average; blue=$M*>10^{11}M_\odot$;
  green=$10^{10}<M*<10^{11} M_\odot$.}
\label{mdotsfr}%
\end{figure}

A proper comparison between AGN activity and host galaxy SFR requires
at least a rough separation between activity in galaxies of different
stellar mass. Santini et al. (2009) and Gruppioni et al. (2015)
provides estimates of the SFR for galaxies separated into two mass
bins, log$M_*>$11 and 10$<$log$M_*<$11. To statistically evaluate the
contribution of AGN wind mass outflow rate into these two galaxy mass
bins we need to statistically associate to each AGN bolometric
luminosity realization a host galaxy stellar mass. This can be done by
associating to the AGN bolometric luminosity a SMBH mass (by assuming
a distribution of Eddington ratios), and then converting the SMBH mass
to a stellar mass. This was done using the results briefly presented
in the previous section and in Shankar et al. (2013, 2016). In
particular, we used the model with $\epsilon=0.06$ and the Eddington
ratio distribution given by eq. 15 of Shankar et al. (2013), and the
SMBH mass - galaxy mass correlation given by eq. 6 of Shankar et
al. (2016), assuming an intrinsic dispersion of 0.4 dex.  The
resulting distributions of AGN wind mass outflow rates (total, and in
the two stellar mass bins given above), have been binned to build AGN
mass outflow rate density functions. $10^8$ realizations have been
randomly chosen in each redshift bin.  Fig. \ref{mdotsfr}, shows the
average mass-loading factor, i.e. the ratio between the resulting AGN
wind outflow rate density and the average SFR density as a function of
the redshift. The average mass-loading factor is between 20\% and 40\%
for the average population and peaks at z$\sim$1. The distributions
are quite different when splitting the galaxy population in low mass
stellar mass galaxies ($10^{10}-10^{11}$ M$_\odot$) and high stellar
mass galaxies ($>10^{11}$ M$_\odot$). Small mass galaxies hosting, on
average, fainter nuclei with energetically fainter AGN winds, are less
affected by AGN winds than larger galaxies, hosting, on average, more
luminous nuclei, with more energetic winds.  The latter galaxies
(stellar masses $>10^{11}$ M$_\odot$) are, on average, strongly
affected by AGN winds at z$\ls2$, where they have $<\eta>\gs1$. The
relative importance of AGN winds reduces at z$>2$ also in massive
systems, remaining however always higher than that in less massive
systems. In this calculation we used the new calibration of the {\it
  intrinsic} SMBH mass-galaxy mass correlation found by Shankar et
al. (2016) to split the average mass loading factor into massive and
less massive systems. Similar conclusions are obtained using the
traditional, {\it biased}, correlation.

We calculated the average mass loading factor by assuming a square
root and a linear scaling between the wind mass outflow rate and the
bolometric luminosity. In the former case the curves shift to slightly
lower redshift (peak redshift between 0.5-1), and lower value of the
average loading factor (10-30\%), while for the latter case the
opposite trend is observed. We also calculate the average mass loading
factor by assuming different normalizations of the $\dot M_{OF} -
L_{bol}$ scaling. To bring the average mass loading factor to $\sim1$
would require a normalization $\sim3$times higher than the dashed line
in Fig.1, left panel, completely inconsistent with the present
data. Changing the normalization within its statistical error does not
change significantly the conclusions described above.

\section{Discussion}

\subsection{AGN wind scaling relations}

As mentioned above, while winds are ubiquitous in AGN, both their
effect on AGN host galaxies and their cumulative effect on galaxy
evolution is poorly understood.  To gain more insight on this topics
we collected wind, AGN and host galaxy data for 94 AGN with
massive winds detected at differente scales and ionization states.

We find a strong correlation between mass outflow rate $\dot M_{OF}$
and AGN bolometric luminosity L$_{bol}$ for both molecular winds
($\dot M_{OF}\propto L_{bol}^{0.76\pm0.06}$) and ionised winds ($\dot
M_{OF}\propto L_{bol}^{1.29\pm0.38}$).  Models implying shocks
expanding into an isothermal sphere (density $\propto R^{-2}$)
predict $\dot M_{OF}\propto L_{bol}^{1/3}$ (Faucher-Giguere \&
Quataert 2012, King \& Pounds 2015). Lapi et al. (2005) finds a $\dot
M_{OF}\propto L_{bol}^{0.5}$ scaling in the case of an isothermal
density profile, if the ratio between the outflow energy and the
energy of the ambient ISM $\Delta E/E$ is kept constant.  Steeper
scalings can be obtained for shocks expanding in a medium where the
density profile is flatter than the isothermal case (Faucher-Giguere
\& Quataert 2012). However, it should be considered that all quoted
models refer to a {\it total} mass outflow rate and do not consider
multi-phases winds, while our determinations concern a particular wind
phase (neutral-molecular or atomic-ionized).

Because the scaling of $\dot M_{OF}$ with L$_{bol}$ of ionised winds
is steeper than that of molecular wind, the ratio between molecular
and ionized mass outflow rate reduces toward high bolometric
luminosity.  For the sources of our sample with molecular gas
estimates we also find that the molecular gas depletion timescale and
the molecular gas fraction, both corrected for the trends with the
redshift and with the distance from the main sequence according to the
recipe of Genzel et al. (2015), are 3-30 times shorter and $\sim10$
times smaller, respectively, than the average of star-forming galaxies
with similar SFR, stellar mass, distance from the main sequence and
redshift. One may speculate then that at high AGN bolometric
luminosity and wind kinetic power, the reduced molecular gas fraction
may be due to the destruction of molecules in the wind, leading to a
large fraction of gas in the atomic/ionised phase. Indeed, models of
molecular shocks at densities $>10^4$ cm$^{-3}$ predicts that the
shock will dissociate H$_2$ and become J type (Draine et al. 1983).
Models of J-type shocks predict efficient reformation of molecules in
the post-shock gas, as well as UV radiation to photo-dissociate CO and
significant HCN formation (Neufeld \& Dalgarno 1989). Interestingly,
where observations sensitive enough to detect HCN do exist,
(e.g. Markarian 231), enhanced HCN broad wings have been revealed
(Aalto et al.  2012, 2015).

We find a strong correlation between $v_{max}$ and the AGN bolometric
luminosity for extended molecular winds (mostly CO winds), ionised
winds and X-ray UFOs.  The scaling of $v_{max}$ with L$_{bol}$ of
molecular+ionised winds is similar to that of UFOs. Both scalings
between bolometric luminosity and maximum velocity are consistent with
$v_{max}$ to the fifth power, similar to the MBH-$\sigma$ scaling. It
is interesting to note that this scaling is similar to that predicted
by Costa et al. (2015) for the case of energy conserving outflows (see
their eqs. 16 and 17). 

It is instructive to compare the latter results with those of samples
of similar (or even bigger) size from Herschel spectroscopy of OH
lines. Relative to CO, the specific characteristic of OH in galaxies
is that lines are likely radiatively (instead of collisionally)
excited, and thus selectively trace a warmer outflow region, closer to
the circum-nuclear source of strong far-IR radiation density. Although
the outflow is not spatially resolved the observed excitation
conditions provide information about the spatial extent of the
outflow, which enables the estimation of the outflow physical
parameters (mass-outflow rate, mechanical power and energy).

Gonzalez-Alfonso et al. (2014) present a detailed analysis of OH signal
in Markarian 231. The blue wing of the absorption detected in the
high-lying 65$\mu$m doublet with $E_{lower}$=290 K, with
high-velocity shifts $>$1000 km/s, indicates that the excited
outflowing gas is generated in a compact and warm (circum)nuclear
region (diameter of a few hundred pc). Aalto et al 2012 and Cicone et
al. 2012 found that the molecular outflow size in Markarian 231
decreases with the critical density (it is smaller for higher CO
transition and for HCN). OH transitions may lie on this
trend. Furthermore, OH outflow velocities and mass-outflow rates are
similar to that derived from CO in the few sources where both CO and
OH winds are detected simultaneously (see the Appendix).

Spoon et al. (2013) analysed the 79$\mu$m and 119$\mu$m OH transitions
in a sample of 24 ULIRGs at z$<$0.26. Veilleux et al. (2013) analysed
the OH 119$\mu$m transition in 43 mergers at z$<$0.3, mostly ULIRGs and
QSOs (six objects are in common with the Spoon et al. 2013 sample).
Both works found that outflows are a common (seen in $>$70\% of the
cases), and that the relative strength of the OH emission component
decreases as the silicate 9.7$\mu$m absorption increases, locating the OH
outflows inside the obscured nuclei. Both authors also found that the
outflow velocity does not correlate with the galaxy SFR while it
correlates with AGN bolometric luminosity, suggesting that, at least
in ULIRGs and QSOs, AGN dominates over star-formation in driving the
outflow (also see Cicone et al. 2014).

More recently, Stone et al. (2016) searched for outflowing OH in 52
local (distance $<$50Mpc) AGN, selected at hard X-ray wavelengths by
Swift BAT.  While OH is detected in absorption in 17 cases, outflows
($v_{84}<$-300km/s) are detected in four cases only (detection rate of
24\%).  Combining this sample with that of Veilleux et al. (2013),
Stone et al. confirm the trend of outflow velocity with AGN bolometric
luminosity. Furthermore, increasing by several order of magnitude the
dynamic range in SFR, a trend of outflow velocity with SFR emerges,
suggesting that at low AGN bolometric luminosities both AGN and
star-formation contribute in driving OH outflows (see also
Gonzalez-Alfonso et al. 2016).

We report in Fig. \ref{pload} the $L_{bol}-v_{max}$ (and
$L_{bol}-v_{84}$) scalings found by Spoon et al. (2013) and Veilleux
et al. (2013) for OH outflows. While the Veilleux et al. scalings
agree quite well with our results, the Spoon et al. (2013) scaling is
somewhat flatter than both the Veilleux et al (2013) scaling and our
scalings. We note that the dynamic range in bolometric luminosity
covered by Spoon et al. (2013) is smaller than that of Veilleux et
al. (2013) and much smaller than ours, and that the uncertainty on the
slope of a correlation depends linearly on the dynamic range.
Unfortunately, Stone et al. (2016) do not publish a best fit
correlation between $v_{84}$ and $L_{bol}$. However, we mark in Fig.
\ref{pload} the loci covered by two groups of Swift BAT AGN with
$42.3<L_{bol}<43.3$ and $43.3<L_{bol}<43.3$ and by the outlier
NGC7479. We note that the Stone et al. results on low luminosity AGN
align reasonably well along the correlation found for our sample, once
allowing for the offset between $v_{84}$ and $v_{max}$.

If the scaling of $v_{max}$ with L$_{bol}$ of molecular+ionised winds
is similar to that of UFOs, than at each given bolometric luminosity
the ratio between UFO maximum velocity and molecular-ionized wind
maximum velocity should be similar and equal to $\sim40-50$. This
implies that the gas mass involved in Galaxy scale outflows should be
1500-2500 times the gas mass involved in nuclear high velocity
winds. This prediction can be verified by measuring in the same
objects both the nuclear and the galaxy scale winds.  So far this has
been possible in three AGN only, Markarian231 (Feruglio et al. 2015),
IRASF11119+13257 (Tombesi et al. 2015) and APM08279 (Feruglio et al.,
in preparation).  In these three cases the ratio between the gas mass
of the nuclear and galaxy scale winds is in the right ballpark, but
more observations of this kind are clearly needed before drawing a
strong conclusion (see also Stern et al. 2016).

\subsection{AGN winds in a cosmological context}

We use the continuity equation to compute the evolution of the SMBH
accretion rate and compare it to the cosmic SFR density. We find that,
the ratio of the {\it average} SMBH mean accretion density and {\it
  average} SFR density is about constant with redshift.  In massive
galaxies the ratio is about constant for constant Eddington ratio
distributions and constant radiative efficiencies, while it decreases
with increasing redshift for an evolving Eddington ratio
distribution and a mass dependent radiative efficiency.  For less
massive galaxies the ratio peaks at z$\sim$0.5-0.7 in all studied
cases.

We evaluate the evolution of the average AGN wind mass-loading factor,
$<\eta>$, the relative importance of AGN winds to deprive
star-formation from its fuel, by convolving the AGN wind - AGN
bolometric luminosity scaling relation with AGN bolometric luminosity
density, and dividing the result for the SFR density. We find that if
$\dot M_{OF} \propto L_{bol}^{0.76}$, as suggested by molecular winds,
$<\eta>$ is between 0.2 and 0.3 for the full galaxy
population. Instead, $<\eta> > 1$ for massive galaxies at z$\ls2$.  A
tentative conclusion is then that AGN winds are, on average, powerful
enough to clean galaxies from their molecular gas (either expelling it
from the galaxy or by destroying the molecules) in massive systems at
least at z$\ls2$. At higher redshifts the uncertainties in both wind
mass outflow rate density and SFRD are today too big to derive solid
conclusios.  Should the scaling between $\dot M_{OF}$ and $L_{bol}$ be
steeper than assumed above, $<\eta>$, would be higher.  The steep rise
of $<\eta>$ between z=1 and z=0.2 for massive galaxies is due to the
equally steep decrease of the SFRD in these systems. We caution that
our results are obtained using a crude splitting of galaxies into two
broad groups, and, of course, the results are sensitive to the
particular galaxy mass threshold adopted for the splitting, in
particular where the trends are steeper.

We remark again that all results presented above are based on
heterogeneous and biased AGN wind samples. In particular, the
relationships between molecular wind properties and AGN/host galaxy
properties are calibrated at low-redshift only. We assume that similar
scalings hold up to z=2-3, which is something that only new deep ALMA,
NOEMA and VLA observations can confirm. Despite these limitations, our
results suggest that AGN wind kinetic energy rate and mass-loading
factor can be large in single systems. They {\it may} still be
important when diluting their effect by accounting for the short AGN
phases compared to the star-formation cosmic timescales. AGN winds may
be the long sought smoking gun of AGN `{\it feedback in action}, in
massive galaxies at z$\ls$2, while at smaller masses other mechanisms
are also likely to be in place (e.g. Peng et al. 2015).

The relationship between AGN winds and SFR does not appear to be
simple, even in the best studied systems with the strongest winds. The
idea that AGN driven winds may simply clean their host galaxies from
dense gas, thus stopping the formation of any new star, is probably an
over-simplistic view of a very complex, non linear process. Winds
inject energy and entropy in the ISM, ionising and heating it
up. Outflowing gas may experience different phases, as our results
suggest (but note that it is not at all clear how dense cold molecular
gas can be involved in these winds, see Ferrara \& Scannapieco 2016). 
A fraction may leave the system and pollute the circum-galactic
medium, but some may rain back into the galaxy disk.  
The gas leaving the galaxy may inject energy, entropy and
metals into the circum-galactic medium (CGM), thus affecting the
cooling of the CGM gas and, in doing so, affecting further gas
accretion into the galaxy. AGN feedback is then likely part of a
complex {\it feeding and feedback cycle}, consistent with a {\it
  strong} form of AGN/galaxy co-evolution. Gas cools down forming
stars and accreting toward the nucleus, giving rise to the growth of
the central SMBH through luminous AGN phases. In turn, the AGN powers
winds that can heat both ISM and the CGM, altering further
star-formation and nuclear gas accretion. The SMBH growth is then
stopped, as well as nuclear activity and winds, until new cold gas
accretes toward the nucleus, so starting a new AGN episode.  In this
cycle winds and feedbacks might be identified with the 'growth
hormone' of galaxies, that regulates and modulates galaxy and BH
growth.

Finally, AGN winds may help in cleaning the way through their host
galaxy, by both removing the gas, and by ionising it
through shocks and high energy radiation, thus allowing ionising
photons from both the AGN and the star-forming regions
to escape in the IGM (Giallongo et al. 2015). This may
contribute to the ionising UV background at high-z, which is
eventually the responsible for re-ionizing the Universe at z=6-8.

\section{Conclusions}

We collected multi-wavelength observations of 94 AGN host galaxies at
various redshifts, characterised by the presence of a wind detected in
a given gas phase. We used these observations to study the scaling
relationships between wind properties, AGN properties and host galaxy
properties. We report the following findings:

\begin{enumerate}
\item
We confirm, over the largest sample available to date, the remarkable
correlation between mass outflow rate and AGN bolometric luminosity
(Fig. 1, left panel). For molecular winds, $\dot M_{OF}\propto
L_{bol}^{0.76\pm0.06}$ while for ionised winds the scaling is $\dot
M_{OF}\propto L_{bol}^{1.29\pm0.38}$ (Table 1). These scalings are
steeper than those predicted by shock models expanding into a medium
with an isothermal density profile. Flatter density profiles may help
in explaining the observed scaling.

\item
The scaling of $\dot M_{OF}$ with L$_{bol}$ is steeper for ionised
winds than for molecular winds, meaning that the ratio between
molecular to ionised mass outflow rates reduces at the highest AGN
bolometric luminosities, i.e. the fraction of outflowing gas in the
ionised phase increases with the bolometric luminosity.

\item
The wind kinetic energy rate $\dot E_{kin}$ is correlated with
L$_{bol}$ (Fig. 1, right panel) for both molecular and ionised
outflows ($\dot E_{kin}/L_{bol}\sim1-10\%$ for molecular winds, $\dot
E_{kin}/L_{bol}\sim0.1-10\%$ for ionised winds).  About half X-ray
absorbers and BAL winds have $\dot E_{kin}/L_{bol}\sim0.1-1\%$ with
another half having $\dot E_{kin}/L_{bol}\sim1-10\%$. A few UFOs may
have $\dot E_{kin}\sim L_{bol}$, although the uncertainties in the
estimate $\dot E_{kin}$ of UFOs is quite large.

\item
$v_{max}$ correlates with the bolometric luminosity for molecular+
  ionized winds and for UFOs (Fig. 2, left panel). Both scalings are
  statistically consistent with each other, implying that, at each
  given bolometric luminosity, the ratio between UFO maximum velocity
  and molecular-ionized wind maximum velocity is $\sim40-50$ and that
  the total gas mass involved in Galaxy scale outflows should be
  1500-2500 times the gas mass involved in nuclear high velocity
  winds.

\item
The momentum load of most molecular winds is $\dot P_{OF}/\dot
P_{AGN}>3$, half have momentum load $>10$, pointing toward molecular
winds observed in the energy conserving phase. About half ionised
winds have momentum load $<1$ with the other half having $\dot
P_{OF}/\dot P_{AGN}>1$ and a few $>10$, suggesting that also several
ionised winds may be energy conserving.  BAL winds and X-ray absorbers
have momentum load in the range 0.01-1. Fast X-ray winds may be
identified with the momentum conserving, semi-relativistic wind phase,
occurring on scales close to the accretion disc. BAL winds share
similar velocities and momentum load of warm absorbers (Fig. 2, right
panel).

\item
Similar to other studies, we found that most molecular winds and the
majority of ionised winds have kinetic power in excess to what would
be predicted if they were driven by SNe, based on the SFR measured in
the AGN host galaxies (Fig. 3, left panel). The straightforward
conclusion is that most powerful winds are AGN driven.

\item 
The AGN wind mass-loading factor, $\eta=\dot M_{OF}$/SFR, is not
strongly correlated with the AGN bolometric luminosity (Fig. 4, left
panel) and is systematically higher than the mass-loading factor of
starburst driven winds at each given galaxy stellar mass (Fig. 4,
right panel).

\item
The depletion timescales and gas fractions of galaxies hosting strong
winds are 3-30 times shorter and $\sim10$ smaller, respectively, than
the average of star-forming galaxies with similar SFR, stellar mass,
distance from the main sequence and redshift (Fig. 6 and 7).

\end{enumerate}

We then attempted to put AGN winds into a broader cosmological
framework to assess the relative importance of AGN winds on the
average SFR, accounting for the short AGN duty cycle. We can summarize
the results as follows:

\begin{enumerate}

\item 
We find that the ratio of the {\it average} SMBH mean accretion
density and {\it average} SFR density is about constant with
redshift. In massive galaxies, the ratio is about constant for
constant Eddington ratio distributions and constant radiative
efficiencies, while it decreases with increasing redshift for an
evolving Eddington ratio distribution and a mass dependent radiative
efficiency.  For less massive galaxies the ratio peaks at
z$\sim$0.5-0.7 in all studied cases (Fig. 11).
\item

Finally, we find that the average AGN wind mass-loading factor,
$<\eta>$ is between 0.2 and 0.3 for the full galaxy population while
$<\eta> > 1$ for massive galaxies at z$\ls2$ (Fig. 12).  A tentative
conclusion is then that AGN winds are, on average, powerful enough to
clean galaxies from their molecular gas (either expelling it from the
galaxy or by destroying the molecules) in massive systems only, and at
z$\ls2$.

\end{enumerate}

AGN wind studies are evolving from childhood to adult age, and much
remains to be understood. The next step is targeting unbiased AGN and
galaxy samples, thus deriving direct information on wind
demography. This is a difficult and time consuming effort which
several on-going programs aim at achieving in the next years (VLT/KMOS
KASHz and KMOS3D surveys, VLT/SINFONI SUPER survey, VLT/SINFONI,
LBT/LUCI WISSH survey, IRAM PHIBBS2 and IBISCO surveys). In
particular, it is crucial to push to high redshift the systematic
study of molecular winds. All this will allow us to measure wind
parameters and SFR in well defined and little biased samples of AGN at
different redshifts, and calculate first the wind mass-loading factor
source by source, and then its average over each redshift range.
Then, we need to assess whether the winds are typically multiphase,
and/or different wind phases are geometrically distinct. Finally we
need to understand the fate of the outflowing gas, whether it remains
in the systems or if it reaches the CGM.
  
{\bf Acknowledgements}

This work was supported by ASI/INAF contract I/009/10/0 and INAF PRIN
2011, 2012 and 2014.  MB acknowledges support from the FP7 Career
Integration Grant ``eEASy'' (CIG 321913).  LZ acknowledges support
from ASI/INAF grant I/037/12/0.  CF acknowledges funding from the
European Union Horizon 2020 research and innovation programme
under the Marie Sklodowska-Curie grant agreement No 664931.  CC
acknowledges funding from the European Union Horizon 2020
research and innovation programme under the Marie Sklodowska-Curie
grant agreement No 664931 and support from Swiss National Science
Foundation Grants PP00P2\_138979 and PP00P2\_166159. FF thanks
Stefano Borgani and Silvano Molendi for useful discussions. We thank
an anonymous referee for comments that helped to improve the
presentation.

\appendix

\section{Source samples}

The source samples used in this paper are detailed below and
  presented in Table B1. As a general rule, we used only AGN for
which there is an estimate (or a robust limit) on the physical size of
the high velocity gas involved in the wind. This is in fact crucial to
obtain an estimate of the outflow rates of mass and kinetic energy.

\subsection{Molecular winds}

The bulk of sample of AGN molecular winds is from the compilation of
Cicone et al. (2014), and includes AGN in ULIRGs and nearby Seyfert
galaxies.  We used additional data from Feruglio et al (2013a,b:
NGC6240), Feruglio et al. (2015: Markarian 231), Krips et al. (2011)
and Garcia-Burillo et al. (2014; NGC1068), Morganti et al. 2013a,b
(IC5066), Alatalo et al. 2011 (NGC1266), Combes et al. (2013;
NGC1433), and Sun et al. (2014; SDSSJ135646.10+102609.0). In all these
cases the winds are traced from high velocity wings observed in
CO(1-0), CO(2-1), CO(3-2). We also added to this sample winds traced
by high velocity OH from Sturm et al. (2011; I13120-5453, I14378-3651
and I17208-0014), and Tombesi et al. (2015; IRASF11119+13257). Two
sources have mass outflow rates computed using both CO and OH
transitions (Markarian 231, Sturm et al. 2011, Gonzalez-Alfonso et
al. 2014, Feruglio et al. 2015; and IRAS 08572+3915, Sturm et
al. 2011, Cicone et al. 2014). In these two cases the CO and OH mass
outflow rates are within 20\%. Several of the sources with detected
high velocity CO have also detected high velocity OH in absorption
(NGC6240, $v_{84}(OH)$=544km/s, $v_{max}(OH)$=1200km/s, Veilleux et
al. 2013; $v_{max}(CO)$=500km/s, Feruglio et al. 2013b; I10565,
$v_{84}(OH)=489$ $v_{max}(OH)=950$, Veilleux et al. 2013,
$v_{max}(CO)=600$km/s Cicone et al. 2014; I23365,
$v_{84}(OH)=604$km/s, $v_{max}(OH)$=1300km/s, Veilleux et al. 2013;
$v_{max}(CO)=$600km/s, Cicone et al. 2014; IC5063,
$v_{84}(OH)=309$km/s, Stone et al. 2016; $v_{max}(CO)=$400km/s,
Morganti et al. 2013). One source with a CO outflow, NGC1068, does not
have strong OH absorption (Stone et al. 2016). It should however be
noted that Herschel samples a much bigger region than that on which
the CO outflow has been detected. In conclusion, mass outflow rates
and velocities of CO outflows and OH outflows seem comparable,
although OH probably traces more compact wind regions than CO(1-0)
(e.g. Spoon et al 2013, Gonzalez-Alfonso 2014). Indeed, Aalto et al
(2012) and Cicone et al. (2012) found that the molecular outflow size
in Markarian 231 decreases with the critical density (it is smaller
for higher CO transition and for HCN). OH transitions may lie on this
trend.

\subsection{Ionised winds}

Ionised winds are from the sample of Harrison et al. (2014, type 2 AGN
at z$<$0.2), Rupke \& Veilleux (2013, local ULIRGs), 
Liu et al. (2013), Cresci et
al. (2015), Brusa et al. (2016), Perna et al. (2015a,b; all X-ray
selected AGN at z$\sim$1.5), Harrison et al. (2012; ULIRGs at
z=2-3.3), Nesvadba et al. (2008; Radio galaxies at z=2.2-2.6), Genzel
et al. (2014; AGN in star-forming galaxies at z=2.1-2.4), Carniani et
al. (2015; luminous type 1 QSOs at z=2.5), Bischetti et al (2017,
hyper-luminous QSOs at z=2.5-3.5).  In all these cases the wind is
traced by high velocity [OIII]$\lambda5007$, H$_\beta$ and/or
H$_\alpha$.  We finally included in the sample the [CII] wind detected
in the z=6.4 QSO SDSSJ1148 by Maiolino et al. (2012) and Cicone et
al. (2015).

\subsection{BAL winds}

We used BAL data only for sources where there is an estimate in the
literature of the size of the ionised gas cloud responsible for the
absorption. In particular, we used QSOs from Borguet et al. (2013),
Moe et al. (2009), Dunn et al. (2010), Korista et al. (2008), Shen et
al. (2011).

\subsection{X-ray  winds}

Most X-ray winds are from the compilation of Tombesi et al. (2013),
which include fast UFOs and slower warm absorbers. We added to this
list Markarian 231 (Feruglio et al. 2015), IRAS11119 (Tombesi et
al. 2015), PDS456 (Nardini et al. 2015) and APM08279 (Chartas et
al. 2009).

\section{Estimates of physical quantities}

\subsection{Outflows quantities}

Different recipes are used by different authors to calculate physical
quantities from observed ones. To make the comparison between
different sources as homogeneous as possible, we recomputed the wind mass
outflow rates and kinetic power rates given in Table B1 by using standard
recipes. 

The wind mass outflow rate is then computed using the
continuity fluid equation:

\begin{equation}
\dot M_{OF}= \Omega ~ R_{OF}^2 ~\rho_{OF}~ v_{max}
\label{fluid}
\end{equation}

where $\rho_{OF}$ is the average mass density of the outflow, $\rm
v_{max}$ is the wind maximum velocity and R$_{OF}$ is the radius at
which the outflow rate is computed, and $\Omega$ is the solid angle
subtended by the outflow.  Assuming a spherical sector, $\rho_{OF}= 3
M_{OF} / \Omega R_{OF}^{3}$, then:

\begin{equation}
\dot M_{OF}=3\times v_{max}\times M_{OF}/R_{OF}  
\end{equation}

Accordingly, $\dot M_{OF}$ represents the instantaneous outflow rate
of the material at the edge R$_{OF}$ (i.e., it is a local estimate)
and it is three times larger than the total outflow mass divided by
the time required to push this mass through a spherical surface of
radius R$_{OF}$.  This estimator does not depend on the solid angle
$\Omega$ subtended by the outflow.  Three key observables then appear
in the definition of the mass outflow rate: $\rm v_{max}$, R$_{OF}$
and $M_{OF}$.

Following Rupke \& Veilleux (2013) we define the maximum wind velocity
as the shift between the velocity peak of broad emission lines and the
systemic velocity plus 2 times the $\sigma$ of the broad gaussian
component ($v_{max}=velocity ~shift_{broad}+2\sigma_{broad}$). We
assigned to each source ($\rm v_{max})$ either using the published
value if it exists, or evaluating it from the published spectra (as in
the case of NGC1068, Krips et al. 2011). Estimating the bulk wind
velocity from the observed velocities is not straightforward, because
the conversion depends by the wind geometry and spatial distribution
of velocities, see the discussion in Liu et al. (2013b). In our
analysis we assumed that the bulk wind velocity is $\sim$ maximum wind
velocity. Other authors suggest that a better proxy for the bulk wind
velocity is W80/1.3 (Liu et al. 2013, Harrison et al. 2014), where W80
is the velocity width of the emission lines at the 80\% of the line
flux.  Several sources of our sample have published W80 and $v_{max}$
velocities. For these sources we find $<W80/v_{max}>=0.96\pm0.16$,
i.e. W80 and $v_{max}$ are very similar, on average within 4\%. We
therefore conclude that by using the recipe W80/1.3 to estimate bulk
wind velocities would produce mass outflow rates and kinetic energy
rates smaller by $\sim35\%$ and by a factor $\sim2.5$ respectively.
Since we use the same recipe for all sources, of course using W80/1.3
instead of $v_{max}$ will not change the results of the trend
analysis.

R$_{OF}$ is taken from the quoted literature. In most cases R$_{OF}$
is taken as the maximum radius up to which high velocity gas is
detected (baseline method). On the other hand, Carniani et al. (2015),
evaluate a size of the ionised wind systematically lower than all
other cases, because they adopt a different astrometric
procedure. This gives rise to mass outflow rates higher than if they
were calculated with the baseline method. For a few sources integral
field spectroscopy observation are not available and the size of the
high velocity gas is estimated using off-centre spectra (Perna et
al. 2015a, Bischetti et al. 2017). This method can detect only gas on
relatively large scales, and therefore the relative mass outflow rates
computed this way probably under-estimate the real rates.

For molecular winds, the mass of the outflowing gas is computed by
converting CO luminosities into H$_2$ masses. This is usually done by
assuming a proper conversion factor $\alpha_{CO}$, which can be a
function of density, metallicity, gas distribution etc. (see Bolatto
et al. 2013 for a review).  We conservatively adopted
$\alpha_{CO}=0.8$ for our sample, mostly made by LIRGs and ULIRGs.

For ionised winds, the mass of the outflowing gas is calculated by
using the following equations (Osterbrock \& Ferland 2006, Carniani et al. 2015):

\begin{equation}
M_{[OIII]} = 4.0\times 10^7 M_\odot ({C \over 10^{O/H}}) ({L_{[OIII]} \over 10^{44}}) ({<n_e> \over 10^3 })^{-1}
\end{equation}
and
\begin{equation}
M_{H\beta} = 7.8\times 10^8 C ({L_{H\beta} \over 10^{44}}) ({<n_e> \over 10^3 })^{-1}
\end{equation}

assuming that the contribution of star-formation to the observed
luminosities of the broad (wind) line emission is negligible, as in
the literature quoted in Table B1, and a gas temperature of
$T=10{^4}$K. The H$_\beta$ emissivity scales nearly linearly with the
inverse of the temperature, so the mass of the outflow would be about
twice for a temperature twice the one we consider here, for each given
H$_\beta$ luminosity. The [OIII] emissivity does not change much with
the temperature for $T$ from a fraction to a few $10^{4}$ K. Both
$M_{[OIII]}$ and $M_{H\beta}$ scale linearly with the average gas
density $<n_e>$. This is often ill-defined, because it can be
estimated only from the ratio of faint [SII] doublet.  Genzel et
al. (2014) use $<n_e>=80$cm$^{-3}$ because this is the mean value in
the disks and centres of star-forming galaxies. Nesvadba et al. (2006,
2008) found $<n_e>=240-570$cm$^{-3}$ and $<n_e>=300-1000$cm$^{-3}$ in
two z$\sim2$ radio galaxies. Harrison et al. (2014) find
$<n_e>=200-1000$cm$^{-3}$ in a sample of low-z AGN. Perna et
al. (2015a) find $<n_e>=120$cm$^{-3}$ in a z=1.5 AGN. In the following
analysis we adopt $<n_e>=200$cm$^{-3}$ for all objects in the
sample. For sources with H$_\beta$ and H$_\alpha$ measurements we used
the gas mass evaluated from these lines. For sources with only [OIII]
measurements we assume that the total ionised gas mass is 3 times the
$M_{[OIII]}$ (this is the average of $M_{H\beta}/M_{[OIII]}$ for the
sources in our composite sample for which estimates of $M_{[OIII]}$
and $M_{H\beta}$ are simultaneously possible).

The largest uncertainty in the evaluation of molecular mass outflow
rates is currently the size of the outflow. This uncertainty will
likely greatly be reduced by future higher resolution ALMA and NOEMA
observations. For ionised outflows, the largest uncertainties are the
size of the outflow, the gas density and the $M_{H\beta}/M_{[OIII]}$
ratio. 

X-ray absorbers wind masses, outflow rates and kinetic power are even
more uncertain than molecular and ionizes gas masses outflow rates and
kinetic power, due to the statistics of X-ray spectra which is usually
not excellent, and due to large systematic uncertainties in the
evaluation on the size of the wind (only lower and upper limit can be
derived from current, low resolution X-ray spectroscopy).  The
situation should greatly improve with the advent of high resolution
micro-calorimeters in the X-ray bands (resolution of several
thousands), which are planned for the Athena mission (Nandra et al. 2013).

\subsection{Host galaxies and AGN quantities}

We collected from literature for our AGN sample AGN bolometric
luminosities, host galaxy star-formation rates, stellar masses and
total molecular gas masses (disk plus outflows), when available.  We
put particular care in searching and reporting star-formation rates,
stellar and gas masses relative to the size of the region interested
by the wind ($R_{OF}$ in Table B1).  AGN bolometric luminosities are
calculated by fitting to the observed UV to optical SEDs AGN + galaxy
templates, and by converting the mid-infrared and or the X-ray
luminosity by using a luminosity dependent bolometric
corrections. When more than one estimate of AGN bolometric luminosity
does exist (e.g. for bright local Seyfert galaxies), we used the one
minimizing the uncertainty due to a) obscuration of the active
nucleus; b) contribution of the host galaxy to the observed
luminosity. An additional source of scatter in AGN bolometric
luminosity is due to AGN variability and the fact that most
observations at different wavelengths are not simultaneous. We
estimate that the total uncertainty on AGN bolometric luminosities can
be the order of half decade. This is still much smaller than the
dynamic range in AGN bolometric luminosity investigated in this paper
(five decades).
  
The SFR reported in Table B1 are from far infrared photometry when
possible. The AGN contribution to the far infrared band is negligible
in most cases, but for the most luminous QSOs. Even in hyper-luminous
QSO, Schneider et al. (2015) and Duras et al. (2017) found that $\ls$
half of the far infrared light is likely produced by dust in the
galaxy disk illuminated by the AGN emission. The infrared computed SFR
is not the instantaneous SFR but rather the conversion from the
observed FIR luminosity produced by dust reprocessing of light from
stars born hundreds of millions of years before. This SFR is therefore
an upper limit to the on going SFR.  Indeed, Davies et al. (2007)
found that the on going SFR in the nuclei of Markarian 231 and NGC1068
is probably very small, because of the small observed Br$\gamma$
equivalent width within 0.1-0.5 kpc from the active nucleus.
  
Stellar masses reported in Table B1 are calculated by modelling
optical-near-infrared galaxy SEDs with galaxy templates or by
converting near infrared luminosities from IFU observations of nearby
AGN host galaxies into stellar masses.

\newpage

\begin{table*}[!ht]
\centering
\caption{AGN wind sample}
\begin{tabular}{lcccccccccl} 
\hline
\footnotesize
Name$^a$ & Redshift & log$L_{bol}$ & log$\dot M_{OF}$ & log$\dot E_{kin}$ & $v_{max}$ & $R_{OF}$ & log$SFR$ & log$M_*$ & log$M_{gas}$  & REF \\
 &  & erg s$^{-1}$ & M$_\odot$/yr  &  erg s$^{-1}$ & km s$^{-1}$ & kpc & M$_\odot$/yr & M$_\odot$ & M$_\odot$  & \\
\hline
\multicolumn{11}{c}{Molecular (CO) winds} \\
\hline
Mrk231     &  0.04217  &  45.7  &      3  &    44.25 &   750 &    0.3  &    1.00  &  9.11  &  8.88   & 1,2,3,4,5  \\
Mrk231     &  0.04217  &  45.7  &   2.84  &    44.21 &   850 &      1  &    2.06  &  9.80  &  9.3    & 1,2,3,4,5  \\
NGC6240    &  0.0248   &  44.8  &   2.70  &     43.6 &   500 &    0.6  &    1.23  &  10.11 &  9.3    & 6,7,8,9  \\
NGC6240    &  0.0248   &  44.8  &   2.08  &    42.79 &   400 &      5  &    2.18  &  11.59 &  9.83   & 6,7,8,9  \\
I08572     &  0.05835  &  45.66 &   3.08  &    44.74 &  1200 &      1  &    1.62  &  11.8  &  9.11   & 10,9  \\
I10565     &  0.04311  &  44.81 &   2.48  &    43.54 &   600 &    1.1  &    1.98  &  11.17 &  9.26   & 10,9,11,12  \\
I23060     &  0.173    &  46.06 &   3.04  &    44.63 &  1100 &      4  &    1.87  &        &  10.39  & 10  \\
I23365     &  0.06448  &  44.67 &   2.23  &    43.29 &   600 &    1.2  &    2.14  &  11.15 &  9.47   & 10,9,11,12  \\
SDSSJ1356  &  0.1238   &  45.1  &   2.54  &    43.44 &   500 &    0.3  &   0.114  &        &  8.48   & 13  \\
NGC1068    &  0.003793 &  43.94 &   2.08  &    42.18 &   200 &    0.1  &   0.204  &  9.30  &  7.8    & 14,15,5  \\
NGC1433    &  0.003589 &  43.11 &   1.03  &    40.89 &   150 &   0.05  &  -0.538  &  9.48  &  7.7    & 16  \\
IC5063     &  0.011    &  44    &   1.34  &    42.05 &   400 &    0.5  &  -0.260  &        &  7.7    & 17,18,19,20  \\
NGC1266    &  0.007318 &  43.3  &   1.11  &    41.73 &   360 &   0.45  &   0.204  &  9.59  &  8.6    & 21,22  \\
\hline
\multicolumn{11}{c}{Molecular (OH) winds} \\     
\hline
I13120     &  0.03076  &  44.84 &   2.11  &    43.48 &   860 &    0.2  &   2.22   &  10.49 &  9.76   & 23,24  \\
I14378     &  0.06764  &  45.43 &   2.87  &    44.51 &  1170 &    0.1  &   1.90   &        &  9.62   & 23,11  \\
I17208     &  0.04281  &  45.11 &   1.95  &    42.59 &   370 &    0.1  &   2.44   &  11.13 &  9.38   & 25, 26,12  \\
I11119     &  0.189    &  45.91 &   2.90  &    44.41 &  1000 &    0.3  &   2.20   &        &  9.95   & 25,26  \\
\hline
\hline
\end{tabular}
\end{table*}

\newpage

\setcounter{table}{0}    
\begin{table*}[!ht]
\caption{AGN wind sample, continue}
\begin{tabular}{lcccccccccl} 
\hline
Name$^a$ & Redshift & log$L_{bol}$ & log$\dot M_{OF}$ & log$\dot E_{kin}$ & $v_{max}$ & $R_{OF}$ & log$SFR$ & log$M_*$ & log$M_{gas}$  & REF \\
 &  & erg s$^{-1}$ & M$_\odot$/yr  & erg s$^{-1}$ & km s$^{-1}$ & kpc & M$_\odot$/yr & M$_\odot$ & M$_\odot$  & \\
\hline
\multicolumn{11}{c}{Ionized ([OIII], H$_\beta$, H$_\alpha$, [CII]) winds} \\
\hline
SDSSJ0945  &  0.1283   &  45.51 &   1.62  &    43.49 &  1511 &    2.7  &   1.91   &        &         & 27  \\
SDSSJ0958  &  0.1092   &     45 &    1.1  &    42.47 &   866 &    2.6  &   1.56   &        &         & 27  \\
SDSSJ1000  &  0.148    &   45.7 &   1.16  &    42.43 &   761 &    4.3  &   1.46   &        &         & 27  \\
SDSSJ10101 &  0.1992   &     46 &   1.82  &    43.69 &  1523 &    3.9  &   2.08   &        &         & 27  \\
SDSSJ10100 &  0.0984   &   45.6 &   1.46  &    43.16 &  1267 &    1.6  &   1.36   &        &         & 27   \\
SDSSJ1100  &  0.1005   &     46 &   1.65  &     43.3 &  1192 &    1.9  &          &        &         & 27 \\
SDSSJ1125  &  0.1669   &   45.2 &   0.74  &    42.63 &  1547 &    2.9  &          &        &         & 27  \\
SDSSJ1130  &  0.1353   &  45.11 &   0.3   &    41.38 &   616 &    2.8  &   1.26   &        &         & 27  \\
SDSSJ1316  &  0.1505   &   45.4 &   1.48  &    43.15 &  1216 &    3.1  &          &        &         & 27  \\
SDSSJ1339  &  0.139    &   44.3 &   0.22  &    41.13 &   505 &    2.5  &          &        &         & 27  \\
SDSSJ1355  &  0.1519   &   45.7 &   0.57  &    41.87 &   797 &    3.5  &          &        &         & 27  \\
SDSSJ1356  &  0.1238   &   45.1 &   1.60  &    43.14 &  1049 &    3.1  &   1.80   &  11.0  &         & 27  \\
SDSSJ1430  &  0.0855   &   45.3 &   1.70  &     43.2 &   999 &    1.8  &   0.85   &        &         & 27   \\
Q1623      &  2.43     &        &   2.45  &    44.18 &  1300 &    1.3  &   1.48   &  10.81 &         & 28   \\
U3-25105   &  2.29     &        &   1.50  &    41.66 &   214 &    1.3  &   1.51   &  10.85 &         & 28  \\
GS3-19791  &  2.22     &   45.6 &   3.23  &    44.18 &   530 &    1.3  &   2.17   &  11.31 &         & 28  \\
D3a-15504  &  2.38     &        &   1.72  &    42.58 &   475 &    1.3  &   1.38   &  11.04 &         & 28  \\
GS3-28008  &  2.29     &   45.9 &   2.34  &     42.8 &   300 &    1.3  &   2.03   &  11.36 &         & 28  \\
COS43206   &  2.1      &        &   2.06  &    42.52 &   300 &    1.3  &   1.64   &  11.4  &         & 28  \\
COS11363   &  2.1      &  46.22 &   2.83  &    44.52 &  1240 &    1.3  &   1.62   &  11.28 &         & 28  \\
SDSSJ1326  &  3.304    &  47.59 &   3.81  &    45.98 &  2160 &      7  &   2.26   &        &         & 29,30   \\
SDSSJ1549  &  2.367    &  47.82 &   3.42  &     45.2 &  1380 &      7  &          &        &         & 29,30  \\
SDSSJ1201  &  3.512    &  47.76 &   3.50  &    45.53 &  1850 &      7  &          &        &         & 29,30  \\
SDSSJ0745  &  3.22     &  47.99 &   3.76  &    45.81 &  1890 &      7  &   3.18   &        &         & 29,30  \\
SDSSJ0900  &  3.297    &  47.91 &   3.52  &    45.77 &  2380 &      7  &   2.90   &        &         & 29,30  \\
LBQS0109   &  2.35     &  47.43 &   2.84  &    44.88 &  1850 &    0.4  &          &        &         & 31  \\
2QZJ0028   &  2.401    &  47.15 &   3.66  &    45.89 &  2300 &    0.7  &   2.00   &        &         & 31  \\
HB8905     &  2.48     &  46.77 &   2.65  &    43.55 &   500 &    1.3  &          &        &         & 31  \\
HE0109     &  2.407    &  47.39 &   3.14  &    44.55 &   900 &    0.4  &   1.70   &        &         & 31  \\
HB8903     &  2.44     &  47.28 &   1.76  &    43.58 &  1450 &    1.9  &   1.95   &        &         & 31  \\
RGJ0302    &  2.239    &  46.34 &   1.48  &    43.17 &  1234 &      8  &   2.93   &        &         & 32  \\
SMMJ0943   &  3.351    &  46.76 &   1.57  &    43.17 &  1124 &     15  &   3.11   &        &         & 32  \\
SMMJ1237   &  2.06     &  46.72 &   1.48  &    43.14 &  1200 &      7  &   2.63   &        &         & 32  \\
SMMJ1636   &  2.385    &  46.28 &   1.44  &    42.99 &  1054 &      7  &   3.15   &        &         & 32  \\
XID2028    &  1.593    &   46.3 &   2.39  &    44.24 &  1500 &     13  &   2.44   &  11.65 & 10.28   & 33,34  \\
XID5321    &  1.47     &   46.3 &   1.84  &    43.93 &  1950 &     11  &   2.36   &  11.7  &         & 35,36  \\
XID5395    &  1.472    &  45.93 &   2.65  &    44.56 &  1600 &    4.3  &   2.57   &  10.89 &         & 37  \\
MIRO20581  &  2.45     &  46.6  &   2.29  &    44.55 &  1900 &    4.8  &   $<2.5$ &  11.28 &         & 38  \\
MRC1138    &  2.2      &   46.6 &   2.39  &     43.7 &   800 &     20  &          &    10  &         & 39  \\
MRC0406    &  2.44     &   46.3 &   3.82  &    45.29 &   960 &    9.3  &          &  8.60  &         & 40 \\
MRC0828    &  2.57     &   46.6 &   3.87  &    45.17 &   800 &      9  &          &        &         & 40  \\
I08572     &  0.05835  &  45.66 &   0.27  &    42.67 &  2817 &      2  &   1.62   &  11.8  &         & 41,10  \\
I10565     &  0.04311  &  44.81 &   0.11  &    41.07 &   535 &      5  &   1.98   &  11.17 &         & 41,10  \\
Mrk231     &  0.04217  &   45.7 &  -0.50  &    40.65 &   665 &      3  &   2.06   &  9.799 &         & 41,2,3,4  \\
SDSSJ0149  &  0.567    &  46.94 &   2.60  &    44.25 &  1191 &    4.1  &   1.82   &  10.8  &         & 42,43,44,45  \\
SDSSJ0210  &   0.54    &  46.16 &   2.62  &    43.62 &   560 &    7.5  &          &  10.2  &         & 42,43,44,45  \\
SDSSJ0319  &  0.626    &  46.44 &   2.32  &    43.76 &   934 &    7.5  &          &  10.6  &         & 42,43,44,45  \\
SDSSJ0321  &  0.643    &  46.51 &   2.30  &    43.75 &   946 &     11  &   1.28   &  11.2  &         & 42,43,44,45  \\
SDSSJ0759  &  0.649    &  47.28 &   2.87  &    44.56 &  1250 &    7.5  &   1.64   &  11.3  &         & 42,43,44,45 \\
SDSSJ0841  &  0.641    &  46.54 &   2.60  &    43.76 &   675 &    6.4  &          &  10.9  &         & 42,43,44,45 \\
SDSSJ0842  &  0.561    &   46.8 &   2.59  &    43.53 &   522 &      9  &   1.18   &  10.1  &         & 42,43,44,45 \\
SDSSJ0858  &  0.454    &  47.23 &   2.79  &    44.24 &   939 &    5.6  &   2.08   &  10.6  &         & 42,43,44,45 \\
SDSSJ1039  &  0.579    &  46.87 &   2.81  &    44.35 &  1046 &    5.8  &   1.57   &  10.6  &         & 42,43,44,45  \\
SDSSJ1040  &  0.486    &  46.19 &   3.16  &    45.19 &  1821 &    7.6  &   1.59   &  10.7  &         & 42,43,44,45 \\
SDSSJ1148  &  6.419    &   47.6 &   3.54  &    45.27 &  1300 &      8  &   3.00   &  11.58 & 10.3    & 46,47  \\
\hline
\hline
\end{tabular}
\end{table*}

\newpage

\setcounter{table}{0}    
\begin{table*}[!ht]
\caption{AGN wind sample, continue}
\begin{tabular}{lcccccccccl} 
\hline
Name$^a$ & Redshift & log$L_{bol}$ & log$\dot M_{OF}$ & log$\dot E_{kin}$ & $v_{max}$ & log$R_{OF}$ & log$SFR$ & log$M_*$ & log$M_{gas}$  & REF \\
 &  & erg s$^{-1}$ & M$_\odot$/yr  &   erg s$^{-1}$ & km s$^{-1}$ & lpc & M$_\odot$/yr & M$_\odot$ & M$_\odot$  & \\
\hline
\multicolumn{11}{c}{BAL winds} \\       
\hline
SDSSJ1106  &  3.038    &   47.2 &   2.59  &     45.9 &  8000 &    0.3  &          &        &         & 48,49  \\
SDSSJ0838  &  2.043    &   47.5 &   2.48  &    45.79 &  8000 &    0.3  &          &        &         & 50  \\
SDSSJ0318  &  1.967    &   47.7 &   2.45  &     45.2 &  4200 &   11.5  &          &        &         & 51  \\
QSO2359    &  0.868    &  47.67 &   1.84  &    43.63 &  1380 &      3  &          &        &         & 52,53  \\
QSO1044    &  0.7      &  46.84 &   2.48  &    45.25 &  4300 &    1.7  &          &        &         & 54  \\
3C191      &  1.956    &  46.57 &   2.49  &    43.99 &  1000 &     28  &          &        &         & 51  \\
FIRST1214  &  0.6952   &  46.43 &   1.44  &    43.55 &  2000 &   0.03  &          &        &         & 55  \\
\hline
\multicolumn{11}{c}{Ultra Fast Outflows} \\
\hline
 &  & erg s$^{-1}$ & M$_\odot$/yr  &   erg s$^{-1}$ & km s$^{-1}$ & cm & M$_\odot$/yr & M$_\odot$ & M$_\odot$  & \\
NGC4151    &  0.003319 &   43.9 &  -2.00  &     42.5 & 3.18$\times10^4$ & 14.6-15.8   &   &        &         & 56  \\
IC4329A    &  0.016054 &   45.1 &  -1.20  &     43.2 & 2.94$\times10^4$ & 15.6-16.5   &   &        &         & 56 \\
Mrk509     &  0.034397 &   45.2 &  -0.75  &    44.15 & 5.19$\times10^4$ & 15.1-16.3   &   &        &         & 56  \\
Mrk509     &  0.034397 &   45.4 &  -0.65  &    44.05 & 4.14$\times10^4$ & 15.3-16.6   &   &        &         & 56  \\
Ark120     &  0.032713 &   45.5 &  -0.70  &    44.65 & 8.61$\times10^4$ & 14.8-17.9   &   &        &         & 56 \\
Mrk79      &  0.022189 &   44.9 &  -0.45  &    43.95 & 2.76$\times10^4$ & 15.3-16.5   &   &        &         & 56  \\
NGC4051    &  0.002336 &   43.3 &  -2.65  &    40.95 & 1.11$\times10^4$ & 14.7-15.9   &   &        &         & 56  \\
Mrk766     &  0.012929 &   44.2 &  -1.80  &     42.5 & 2.46$\times10^4$ & 13.8-17.2   &   &        &         & 56  \\
Mrk766     &  0.012929 &   44.4 &  -1.70  &     42.6 & 2.64$\times10^4$ & 13.7-16.1   &   &        &         & 56  \\
Mrk841     &  0.036422 &   44.9 &  -0.90  &       43 & 1.65$\times10^4$ & 15.8-18     &   &        &         & 56  \\
1H0419-577 &  0.104    &   45.6 &   0.50  &     44.7 & 2.37$\times10^4$ & 16.3-17.9   &   &        &         & 56  \\
Mrk290     &  0.029577 &   44.6 &  -0.55  &    44.35 & 4.89$\times10^4$ & 14.8-16.7   &   &        &         & 56  \\
Mrk205     &  0.070846 &   45.2 &  -0.20  &     44.2 &    3$\times10^4$ & 16.1-16.2   &   &        &         & 56  \\
PG1211+143 &  0.0809   &   45.3 &   0.50  &     45.3 & 4.53$\times10^4$ & 15.3-18.5   &   &        &         & 56  \\
MCG-5-23-16&  0.008486 &   44.5 &  -1.10  &     43.5 & 3.48$\times10^4$ &   15-16.6   &   &        &         & 56  \\
NGC4507    &  0.011801 &   44.4 &  -2.15  &     42.9 & 5.97$\times10^4$ & 13.3-16.9   &   &        &         & 56  \\
Mrk231     &  0.04217  &   45.7 &   0.15  &    43.95 & 2.01$\times10^4$ & 15.7-16.5   &   &        &         & 1  \\
PDS456     &  0.184    &     47 &   1.30  &     46.3 & 1.05$\times10^5$ & 16.2-16.2   &   &        &         & 57  \\
I11119     &  0.189    &   45.9 &   0.50  &     45.3 & 7.65$\times10^4$ & 14.2-15.9   &   &        &         & 58  \\
APM08279   &  3.91     &  47.45 &   1.05  &    46.85 & 1.08$\times10^5$ &   14-16     &   &        &         & 59  \\
\hline
\multicolumn{11}{c}{Warm absorbers} \\    
\hline
NGC3783    &  0.009730 &   44.6 &  -0.10  &    42.35 &  3000    & 17-19.1     &   &        &         & 56  \\
NGC3783    &  0.009730 &   44.4 &  -0.55  &    41.55 &  2100    & 17.3-18.1   &   &        &         & 56  \\
NGC3783    &  0.009730 &   44.5 &  -0.60  &    41.55 &  2100    & 17.3-18.1   &   &        &         & 56  \\
NGC3516    &  0.008836 &   44.8 &  -1.00  &       41 &  1800    & 17.1-17.1   &   &        &         & 56  \\
NGC3516    &  0.008836 &   44.7 &  -0.95  &     41.3 &  2400    & 16.8-16.6   &   &        &         & 56  \\
NGC3516    &  0.008836 &   44.6 &  -1.00  &     41.5 &  3300    & 16.6-16.7   &   &        &         & 56  \\
NGC3516    &  0.008836 &   44.7 &  -1.05  &     41.6 &  3900    & 16.4-16.7   &   &        &         & 56  \\  
Mrk279     &  0.030451 &   45.1 &  -0.60  &     41.5 &  2100    & 17.3-17.9   &   &        &         & 56 \\
ESO323-G77 &  0.015014 &     45 &  -0.35  &    42.25 &  3600    & 16.7-17     &   &        &         & 56 \\
\hline
\hline
\end{tabular}
$^a$Short name. See quoted bibliography for full names.

References:
1=Feruglio et al. 2015; 2=Londsdale et al. 2003; 3=Davies et al. 2004, 4= Veilleux et al. 2009, 5=Davies et al. 2007;
6=Feruglio et al. 2013a; 7=Tacconi et al. 1999; 8- Engel et al. 2010; 9=Howell et al. 2010;
10=Cicone et al. 2014; 11=Dasyra et al. 2006; 12=Downes \& Solomon 1998; 
13=Sun et al. 2014; 
14=Garcia-Burrillo et al. 2014; 15=Krips et al. 2012;
16=Combes et al. 2013;
17=Morganti et al. 1998; 18=Morganti et al. 2013; 19=Woo \& Urry 2002; 20=Malizia et al. 2007;
21=Alatalo et al. 2011; 22=Alatalo et al. 2014;
23=Sturm et al. 2011; 24=da Cunha et al. 2010;
25=Veilleux et al. 2013; 26=Xia et al. 2012;
27=Harrison et al. 2014;
28=Genzel et al. 2014, assuming $H_\alpha/H_\beta=2.9$, extinction corrected;
29=Bischetti et al. 2017; 30=Duras et al. 2017;
31=Carniani et al. 2015;
32=Harrison et al. 2012;
33=Cresci et al. 2015, extinction corrected; 34=Brusa et al. 2015b
35=Brusa et al. 2015a; 36=Perna et al. 2015a, extinction corrected; 
37=Brusa et al. 2016;
38=Perna et al. 2015b;
39=Nesvadba et al. 2006,assuming $H_\alpha/H_\beta=2.9$, extinction corrected;
40=Nesvadba et al. 2008;
41=Rupke \& Veilleux 2013;
42=Liu et al. 2013a; 43=Liu et al. 2013b, extinction corrected; 44=Wylezalek et al. 2016; 45=Reyes et al. 2008;
46=Maiolino et al. 2012; 47=Cicone et al. 2015 [CII] wind;
48=Borguet et al. 2013; 49=Bandara et al. 2009;
50=Moe et al. 2009;
51=Dunn et al. 2010;
52=Korista et al. 2008; 53=Bautista et al. 2010;
54=de Kool et al. 2002; 
55=Shen et al. 2011;
56=Tombesi et al. 2012;
57=Nardini et al. 2015;
58=Tombesi et al. 2015; 
59=Chartas et al. 2009.
\normalsize

\end{table*}

\newpage

\end{document}